\documentclass{article}

\usepackage[english]{babel}

\usepackage[letterpaper,top=2cm,bottom=2cm,left=3cm,right=3cm,marginparwidth=1.75cm]{geometry}

\usepackage{amssymb,amsmath,latexsym}
\usepackage{graphicx}
\usepackage[colorlinks=true, allcolors=blue]{hyperref}

\title{Wave propagation and transformation in the frame of magnetohydrodynamics with a vortex electric field}

\author{Ilya A. Kondratyev$^{1}$\footnote{mrkondratyev95@gmail.com}, Sergey G. Moiseenko$^{1,2}$\footnote{moiseenko@iki.rssi.ru} and Gennady S. Bisnovatyi-Kogan$^{1,3}$\footnote{gkogan@iki.rssi.ru}\\
	{\it $^1$Space Research Institute RAS}\\
	{\it Profsoyuznaya st. 84/32, Moscow 117997} \\
	{\it $^2$National Reserach University} \\
	{\it $"$Higher School of Economics$"$, Physics dept.}\\
	{\it Staraya Basmannaya st. 21/4k5, Moscow 105066}\\
	{\it $^3$National Research Nuclear University $"$MEPhI$"$}\\
	{\it Kashirskoe hw. 31, Moscow 115522} 
}

\date{}

\begin{document}

	\maketitle
	
	\begin{abstract}

Bright transient objects in different wave bands have been discovered  in recent years. To explain these short (from ms to s), and very powerful events different models, galactic and extragalactic, have been considered. One of popular model is based on the suggestion of transformation of the magnetized plasma blob, presumably a MHD shock wave, moving with  relativistic speed, into electromagnetic  pulse after collision with some obstacle,  or at entering in media with different properties.  Such transformation  cannot be described in the frame of classical MHD equations, which don’t contain the displacement current, and don’t include such transformation. Updating of the classic MHD equations by using of full set of Maxwell equation with displacement current, made recently by several authors,  gave a possibility to describe hydrodynamic phenomena, together with electromagnetic ones by the same set of equations. Here we apply these equations for study of propagation of different waves with account of displacement current, and of transformation between different kind of waves in the nonuniform media. We derive a dispersion equation describing a propagation of a weak linear wave in a uniform magnetized plasma, and consider in details the cases with perpendicular and parallel motion  to the  magnetic field. It contains MHD, HD, Alfven and EM waves in the limiting cases, and some new types of behaviour in a general situation. We consider a plasma with zero viscosity and  heat conductivity, with a finite scalar electric conductivity. We have shown, that transformation  of MHD into EM wave may happen only in a highly magnetized plasma, and have found the region of parameters, where no waves exist, and only damping static perturbations may be present.

\end{abstract}
	

\section{Introduction}
	 
Approximate description of behaviour of magnetized media was suggested by H.Alfven in 1942, and called now as magnetohydrodynamics (MHD) \cite{alfven}. It does not take into account a displacement electrical current, and therefore it is good only for sufficiently slow processes \cite{alfven,kulik,lleldin}.
In practice almost all processes with conducting magnetized  fluid on Earth satisfy these conditions. Presently laboratory experiments occurs in conditions, when displacement currents and generation of electromagnetic waves (EM) become important. Namely, it happens in the experiments on the collisions of rapid plasma clouds with magnetic wall \cite{sbk21,kns23,rig}. 

In  astrophysical objects, e.g. X-ray pulsars, AGN, GRB, where strongly magnetized objects, often moving with relativistic speeds are expected, it is necessary to describe these events by equations of relativistic MHD, or of classic MHD with account of displacement currents, for having a possibility to describe transformations between MHD and EM waves, which are present is space.   
		
As a result of plasma penetration through the magnetic wall in nonrelativistic laboratory experiments we expect a formation of MHD shock waves, as well as of electromagnetic ones. To study these processes we need equations following from hydrodynamic and Maxwell equations, which in the limiting situations describe MHD and EM waves, and are valid for the general case, when both phenomena are present.
		
Here we consider a set of equations following from classical HD and Maxwell equation, without neglecting a displacement current, needed for  a formation of EM waves, calling them as electromagnetohydrodynamic  (MHDE) equations. 

Full Maxwell Hydrodynamics equations (FMHD), equivalent to our MHDE, had been introduced in the paper by \cite{mcgre}, where by account of the displacement current the authors had overcame the problem of very large classic Alfven velocity in a rarefied plasma. They considered in detail numerical problems, solved by the presence of displacement current.
We discuss here the transformations between the linear waves, basing on  solutions of the dispersion equation at different parameters. We use here the name MHDE, stressing the importance of the correct consideration of the electric field, while for mathematicians the name FMHD could be more favourable. Numerical investigation of waves in noncollisional plasma with account of displacement current had been performed in \cite{bret}.
 
Analysis of linear waves in the magnetized two-fluid, two-temperature plasma, in 2-D MHD, with account of displacement current, had been done by \cite{liang} for non-relativistic motion. Contrary, we consider equation of classic MHD with displacement current, similar to \cite{mcgre}, and consider finite value of electrical conductivity. Its variability qualitatively change the behavior of waves, their propagation and transformation. \cite{liang}  considered the ideal plasma with infinite conductivity. Similar approach was used earlier in \cite{stas} for extended study of small-scale alfvenic structure in the Aurora. That exclude the propagation of electromagnetic waves in their consideration, what one of the main goals of our analysis. 

Numerical calculations in MHD with finite conductivity in presence of displacement current had been performed by \cite{seyfer} in application to dense Z-pinches.

We find a dispersion equation describing a propagation of a weak linear wave in a uniform magnetized plasma, in the external constant magnetic field. Waves, propagating perpendicular to external magnetic field, in limiting cases belong to  MHD, HD or EM types. For general choice of parameters these equation describes waves having properties of all simple waves. Waves, propagating along the external  magnetic field in the limiting cases describe HD, EM and Alfven waves. The last two types of waves are described analytically by a general formula which explicitly indicates a possibility of  mutual transformation between them. In general consideration a plasma is consider at zero viscosity and heat conductivity, but with a finite electric conductivity of a scalar type.

Most works in relativistic MHD are devoted to numerical investigations of relativistic jets, see e.g. \cite{buc,leis,matt}. In the case of accretion into magnetized neutron star in  X-ray sources the falling velocity in the gravitational field does not exceed the surface free-fall velocity which is $\lesssim 0.5 \, c$. In collision with a NS magnetosphere this velocity is usually $\lesssim 0.1\, c$. Our results could be appropriate for accreting magnetized NS and white dwarfs, for collisions of stellar winds with magnetized interstellar clouds, and to all laboratory experiments, where a rapid plasma beam is colliding with a magnetic wall \cite{sbk21,kns23,rig}. Classical MHDE is applicable also for numerical simulations of the very origin of jets, accelerated by the radiation or magnetic forces from the accretion disk surface.

\section{MHDE equations}
	
Consider a plasma flow with velocity field ${\bf u(x)}$, which is moving in a region with a variable magnetic field ${\bf H(x)}$ and  density $\rho({\bf x})$. It is supposed, that plasma has a scalar conductivity $\sigma$. By neglecting the Hall effect and thermodiffusion \cite{chap90,brag57}, the Maxwell equations and the Ohm's law can be written as \cite{syr57}
	\begin{equation}
	\begin{gathered}
	\frac{1}{c}\frac{\partial{\bf H}}{\partial t}=-{\rm rot}{\bf E},\quad
	\frac{1}{c}\frac{\partial{\bf E}}{\partial t}+\frac{4\pi}{c}{\bf j}={\rm rot}{\bf H}, \quad {\rm div}{\bf H}=0,\\
	  {\rm div}{\bf E} = 4\pi \rho_e, \quad {\bf j}=\sigma({\bf E}+\frac{1}{c}{\bf [u\times H]}),
	\end{gathered}
	\label{maxwellohm}
	\end{equation}
where  ${\bf E}$ is an electric field, ${\bf j}$ is an electric current density, and $\rho_e$ is an electric charge density, which is set to zero hereafter.	
	
Expressing ${\bf E}$ from the Ohm's law, inserting it into the first equation in \eqref{maxwellohm}, and  expressing ${\bf j}$ from the second equation in \eqref{maxwellohm}, we obtain
	\begin{equation}
	\frac{\partial{\bf H}}{\partial t}={\rm rot}{\bf [u\times
		{\bf H}]}-{\rm rot}\left(\frac{c\,{\bf j}}{\sigma}\right), \quad
	{\bf j} = \frac{c}{4\pi} {\rm rot} {\bf H}-\frac{1}{4\pi}\frac{\partial{\bf E}}{\partial t}.
	\label{fieldandcurrent}
	\end{equation}
By eliminating ${\bf j}$, we get from \eqref{fieldandcurrent}
	\begin{eqnarray}
	\frac{\partial{\bf H}}{\partial t}={\rm rot}{\bf [u\times H]}
	-{\rm rot}(\nu_m\,{\rm rot}{\bf H})
	+\frac{c}{4\pi}
	\frac{\partial}
	{\partial t}\left({\rm rot}\frac{{\bf E}}{\sigma}\right), \nonumber\\
	{{\mbox{where}}}\qquad \nu_m=\frac{c^2}{4\pi\sigma},
	\qquad
	{\rm rot}\frac{{\bf E}}{\sigma}=\frac{{\rm rot}{\bf E}}{\sigma}+{\rm grad}\left(\frac{1}{\sigma}\right)\times {\bf E}.
	\label{Bfield0}
	\end{eqnarray}
By using the first equation from \eqref{maxwellohm}, and assuming $\sigma$ to be independent on time, we obtain the equation for the magnetic field in the form
	
	\begin{equation}
	\begin{gathered}
	\frac{\partial{\bf H}}{\partial t}={\rm rot}{\bf [u\times H]}
	-{\rm rot}(\nu_m\,{\rm rot}{\bf H})
	-\frac{\nu_m}{c^2}\frac{\partial^2{\bf H}}{\partial t^2}+
	\frac{{\rm grad}(\nu_m)}{c}\times \frac{\partial{\bf E}}{\partial t}, \\
	{\bf{{\mbox{where}}}} \qquad {\rm rot}(\nu_m\,{\rm rot}{\bf
		H})=-\nu_m \Delta {\bf H}+{\rm grad}(\nu_m)\times {\rm rot}
	{\bf H}.\qquad
	\end{gathered}
	\label{Bfield}
	\end{equation}
In a general case, this equation cannot be solved without  additional description of the behaviour of the electric field ${\bf E}$. In addition to the Maxwell equation for ${\bf E}$ (first equation in \eqref{maxwellohm}), we need to add the equations for the velocity ${\bf u}$, matter density $\rho$ and pressure $P$, which in the absence of viscosity and thermal conduction have the form \cite{kulik}
	
	\begin{eqnarray}
	\frac{\partial\rho}{\partial t}+{\rm div}(\rho{\bf u})=0, \label{mhd0a} \\
	\frac{\partial{\bf u}}{\partial t}+ ({\bf u\nabla)u}  =
	-\frac{1}{\rho}{\bf \nabla}P+\frac{1}{c\rho}[{\bf{ j \times H}}],\label{mhd0b} \\
	\rho T[\frac{\partial{s}}{\partial t}+ ({\bf u\nabla)}s]=
	\frac{{\bf j}^2}{\sigma}.\qquad
	\label{mhd0}
	\end{eqnarray}
By using the expression for ${\bf j}$ from the second relation in	\eqref{fieldandcurrent}, we obtain the equations for the fluid motion in the fields ${\bf H}$ and ${\bf E}$
	\begin{eqnarray}
	\label{mhdfluida}
	\frac{\partial\rho}{\partial t}+{\rm div}(\rho{\bf u})=0, \qquad\qquad \qquad \qquad\\
	\frac{\partial{\bf u}}{\partial t}+ ({\bf u\nabla)u}  =
	-\frac{1}{\rho}{\bf \nabla} \left(P+\frac{H^2}{8\pi} \right)+
	\frac{({\bf H\nabla)H}}{4\pi\rho}-\frac{1}{4\pi c \rho}
	\left[ \frac{\partial{\bf E}}{\partial t}\times {\bf H}\right],
	\label{mhdfluidb}\\
	\rho T[\frac{\partial{s}}{\partial t}+ ({\bf u\nabla)}s]=
	\frac{\nu_m}{4\pi}({\rm rot}{\bf H})^2-\frac{2\nu_m}{4\pi c}
	\left({\rm rot}{\bf H}\cdot  \frac{\partial{\bf E}}{\partial t}\right)+\frac{\nu_m}{4\pi c^2}\left(\frac{\partial{\bf E}}{\partial t}\right)^2.
	\label{mhdfluid}
	\end{eqnarray}
Here $s$ is a specific entropy, $ds=d\varepsilon+PdV$, where $V$ is a specific volume $V=1/\rho$, and $T$ is a matter temperature.

The equations of MHDE describe a one-component gas or liquid in presence of general type of electromagnetic field, described by set of Maxwell equations, including the displacement current. In this approach the collisions and interaction in the media should be effective to establish the state close to a local thermodynamic equilibrium. The matter is treated in a classic way, so these equations can be used only, when  matter velocity and sound speed are considerably less than the light velocity, say $v,c_s\lesssim 0.3\, c$. The equations of relativistic magneto-hydrodynamics (RMHD) \cite{buc} differ from MHDE only by removal of the restriction on the matter and sound wave velocities, which may be arbitrary close to the speed of light. The RMHD equations are used also in the ultra-relativistic limit, at $$\gamma = \frac{1}{\sqrt{1-\frac{v^2}{c^2}}}\rightarrow \infty.$$   
The RMHD equations have a very complicated form, and used mainly for numerical investigation of relativistic jets, observed in quasars (QSO) and active galactic nuclei (AGN).  

For constant values of $\sigma$ and $\nu_m$ in space the equation \eqref{Bfield} is written as

	\begin{eqnarray}
	\label{mhdfieldB}
	\frac{\partial{\bf H}}{\partial t}={\rm rot}{\bf (u\times H)}
	+\nu_m \Delta{\bf H}-\frac{\nu_m}{c^2}\frac{\partial^2{\bf H}}{\partial t^2}, 
	\end{eqnarray}
In the large conductivity limit ($\nu_m \ll 1$) we obtain from \eqref{mhdfieldB} the equation for the ideal MHD  \cite{syr57,lleldin}, while in the low conductivity limit ($\nu_m \gg 1$) this equation describes  electromagnetic waves in a dielectric media.

\section{Linear wave propagation in a uniform plasma with a constant magnetic field and electrical conductivity}
		
The MHDE system, includes equations \eqref{mhdfluida},\eqref{mhdfluidb},\eqref{mhdfluid}, equation \eqref{Bfield} for the magnetic field $\bf H$, and first equation from \eqref{maxwellohm} for the electric field $\bf E$. Equations for linear waves, formed in the uniform plasma at rest, with a constant magnetic field, are obtained using a linearized form of these equations. Arbitrary angle between directions of wave propagation, and uniform magnetic field is used. Consider a uniform plasma at rest, at constant values of $\sigma$, with parameters
	\begin{equation}
	{\bf u}=0,\quad \rho=\rho_0,\quad P=P_0,\quad H_{x 0}=H_{0}={\rm const},\quad H_{y 0}=H_{z0}=E_{z0}=0,
	\label{backgr}
	\end{equation}
where coordinates $"x"$ and $"y"$ correspond to directions, which are parallel and perpendicular to the field $H_0$, respectively, $z$-coordinate is perpendicular to both of them.

Consider waves in this static model which, by appropriate choice of the coordinate system, always  propagate in the direction, perpendicular to ”z” axis. In this situation the perturbed values $u_z=H_z=E_x=E_y=0$ remain zero. Other perturbed values depend on time $t$, and two space coordinates $x,y$. For small perturbations variables we use following notations

	\begin{eqnarray}
	u_x,\,\,u_y,\,\, q=\rho-\rho_0,\,\, p=P-P_0,\,\, h_x=H-H_0, \,\,
	h_y, \,\, E\equiv E_z.
	\label{perturb}
	\end{eqnarray}
For linear perturbations, the entropy perturbation has a second order of smallness, according to \eqref{mhdfluid}, and in linear approximation perturbations are adiabatic. Other MHDE equations linearized around the uniform medium with constant magnetic field $H_0$ along $x$ axis, for 2-D perturbations are written in the form

 \begin{eqnarray}
	\label{linearmhd1}
	\frac{1}{\rho_0}\frac{\partial q}{\partial t}+\frac{\partial u_x}{\partial x}+\frac{\partial u_y}{\partial y}=0, \qquad\qquad\qquad\\
	\frac{\partial h_x}{\partial t}=-H_0\frac{\partial u_y}{\partial y} + \nu_m\left( \frac{\partial^2 h_x}{\partial x^2}+\frac{\partial^2 h_x}{\partial y^2}-\frac{1}{c^2}\frac{\partial^2 h_x}{\partial t^2} \right), \\
	\frac{\partial h_y}{\partial t}=H_0\frac{\partial u_y}{\partial x} + \nu_m\left( \frac{\partial^2 h_y}{\partial x^2}+\frac{\partial^2 h_y}{\partial y^2}-\frac{1}{c^2}\frac{\partial^2 h_y}{\partial t^2} \right), \\
	\frac{\partial u_x}{\partial t} + \frac{c_s^2}{\rho_0}\frac{\partial q}{\partial x}  = 0, \qquad\qquad\qquad \\
	\frac{\partial u_y}{\partial t} + \frac{c_s^2}{\rho_0}\frac{\partial q}{\partial y} + \frac{u_A^2}{H_0}\left(\frac{\partial h_x}{\partial y} -\frac{\partial h_y}{\partial x} + \frac{1}{c}\frac{\partial E}{\partial t}\right)  = 0, \\
	\frac{1}{c}\frac{\partial h_x}{\partial t} + \frac{\partial E}{\partial y} = 0. \qquad\qquad\qquad
	\label{linearmhd6}
	\end{eqnarray}
  
In the momentum equations we use for the adiabatic perturbations of pressure $p =\left(\frac{\partial P}{\partial \rho}\right)_{ent}q=c_s^2q $, with $c_s^2$ as a squared sound speed, the standard MHD Alfven speed $u_A$ is determined as $u_A^2=\frac{H_0^2}{4\pi\rho_0}$.
		
The solution of this system is looked for in the exponential presentation, proportional to $\sim\exp(ik_x x + ik_y y - i\omega t)$.   Using this presentation for variables with constant coefficients $q,\,\,h_x,\,\,h_y,\,\,u_x,\,\,u_y,\,\, E$ in the system \eqref{linearmhd1}-\eqref{linearmhd6}, we obtain the following algebraic system for constant coefficients 
	\begin{eqnarray}
	\label{lineareq}
	\omega q = \rho_0(k_x u_x + k_y u_y), \nonumber\\
 \omega h_x =H_0 k_y u_y - i\nu_m(k_x^2+k_y^2-\frac{\omega^2}{c^2}) h_x, \nonumber \\
  \omega h_y = -H_0 k_x u_y - i\nu_m(k_x^2+k_y^2-\frac{\omega^2}{c^2}) h_y,\nonumber\\
  \omega u_x = k_x\frac{c_s^2}{\rho_0} q, \\
   \omega u_y =k_y\frac{c_s^2}{\rho_0} q + \frac{u_A^2}{H_0}( k_y h_x - k_x h_y - \frac{\omega}{c} E), \nonumber\\
   \omega h_x = c k_y E \nonumber.
	\end{eqnarray}

The dispersion equations for linear waves $\omega(k_x,k_y)$   defined by equating to zero the determinant of this homogeneous linear algebraic system, is written as 
	\begin{eqnarray}
	(k^2c^2-\omega^2)(k^2c_s^2-\omega^2)-\nonumber \\i\frac{\omega c^2}{\nu_m}\left[k^2(c_s^2+u_A^2)-\omega^2\left(1+\frac{u_A^2}{c^2}\right)+k_x^2c_s^2\frac{u_A^2}{c^2}\left(1-\frac{k^2c^2}{\omega^2}\right)\right]=0, \label{disprel}\\
 \omega+i\nu_m\left(k^2-\frac{\omega^2}{c^2}\right)=0.\qquad (k^2=k_x^2 + k_y^2)
	\label{disprel_1}
	\end{eqnarray}

For some particular cases  equation similar to Eq.\eqref{disprel} was obtained by \cite{mcgre}.

\section{ Waves propagating across the magnetic field}

Waves, propagating in the direction perpendicular to the magnetic field have $k_x=0,\,\, k_y^2=k^2$, and from  \eqref{disprel} we obtain the dispersion equation for these waves in the form 

\begin{equation}
    (k^2c^2-\omega^2)(k^2c_s^2-\omega^2)-i\frac{\omega c^2}{\nu_m}\left[k^2(c_s^2+u_A^2)-\omega^2\left(1+\frac{u_A^2}{c^2}\right)\right]=0.
    \label{disprel_trans}
\end{equation}

\subsection{ Limiting cases }
	
In a case of almost ideal plasma $(\nu_m \ll  c^2/ \omega )$ we obtain from \eqref{disprel_trans} in zero approximation $(\nu_m=0)$ 
 
 \begin{equation}
        \omega_{f0}^2 = k^2\frac{u_A^2+c_s^2}{1+\frac{u_A^2}{c^2}}, \quad \omega_s = 0,
	\label{fast_slow}
	\end{equation}
for fast (f) and slow (s) MHD waves. Here a zero frequency (static perturbation) of the slow MHD wave corresponds to the limit of slow MHD wave for propagation across the magnetic field  \cite{kulik,lleldin}. The frequency of the fast MHD wave here differs from the usual MHD because of effects due to electric field. In vacuum case with $u_A \gg c$, this wave transforms into electromagnetic one, while in the opposite case one can obtain a usual value for MHD, where displacement current is neglected.
  
In presence of low dissipation, the static perturbation and fast MHD wave are damping. With a small addition, $\omega_f$ in \eqref{fast_slow} is written in the form
	\begin{equation}
	\omega_f=k c_f+i\gamma_f,\quad \omega_f^2=k^2c_f^2 +2ikc_f\gamma_f,\quad c_f=\frac{\omega_{f0}}{k},
	\label{mhddamp0}
	\end{equation}
giving the following expression for the value of damping increment 
 
	\begin{equation}
	\gamma_f = -\nu_m \frac{k^2 u_A^2}{2(c_s^2+u_A^2)}\left(\frac{c^2-c_s^2}{c^2+u_A^2}\right)^2.
	\label{fast_damp}
	\end{equation}
It follows from the latter equation, that in limiting cases of very dense $(c_s\rightarrow c)$, and very rarefied $(u_A\gg c)$ medium, the wave damping goes to zero, even at finite $\nu_m$. In limiting cases of very dense $(c_s\rightarrow c)$, and very rarefied $(u_A\gg c)$ medium, the wave damping does not occur even with finite $\nu_m$.
	
For weak damping of the static perturbation, we obtain the increment
	\begin{equation}
	\gamma_S = -\nu_m \frac{k_y^2 c_s^2}{c_s^2+u_A^2}.
	\label{slowdamp_1}
	\end{equation}
 
The equation \eqref{disprel_trans} should have four non-trivial roots in all asymptotics. So that, another option for the static perturbation is possible, namely the one, where it tends to drastically decay, when approaching the limit of ideal MHD. It can be considered as a second slow mode. By imposing the value $\omega_{S*} = 0 + i\gamma_{S*}$ in equation \eqref{disprel_trans} and assuming, that $\gamma_{S*}$ is much larger, than $kc, kc_s, ku_A$, one can obtain the limit of a strong damping for the static perturbation in a highly conducting medium:
    \begin{equation}
	\gamma_{S*} = -\frac{c^2 + u_A^2}{\nu_m}.
	\label{slowdamp_2}
	\end{equation}
The behaviour of static perturbation at arbitrary $\nu_m$ is plotted in Figs.\ref{fig:case1r}, \ref{fig:case1i}.  
 
In the opposite case of plasma with very low conductivity 
$(\nu_m \gg  c^2/ \omega )$ from equation \eqref{disprel_trans} one can write solution in zero approximation $(\nu_m=\infty)$ solutions
	\begin{equation}
 \omega_{em}^2 = k^2 c^2, \qquad \omega_{s}^2 = k^2c_s^2.	
	\label{em_sound}
	\end{equation}
corresponding to electromagnetic (em) and sound waves (s).

With small additions in \eqref{em_sound}
	\begin{equation}
	\begin{gathered}
	\omega_{em}=kc+i\gamma_{em}, \quad \omega_{em}^2=k^2c^2+2ikc\gamma_{em},\\
	\quad \omega_s=kc_s+i\gamma_s,\quad
	\omega_s^2=k^2c_s^2+2ikc_s\gamma_s,
	\end{gathered}
	\label{additionem}
	\end{equation}
we obtain, after inserting \eqref{additionem} into \eqref{disprel}:
	\begin{equation}
	\gamma_{em}=-\frac{c^2}{2\nu_m},
	\quad  \gamma_s=-\frac{u_A^2}{2\nu_m}
	\label{decremem}
	\end{equation}
Solution $\omega_{em}$ corresponds to an electromagnetic wave with damping. It damps weakly, i.e., $\omega_{em} \gg \gamma_{em}$,  when strong Ohmic dissipation is presented with $2k\nu_m \gg c$, $\sigma\ll \frac{c}{2\pi k}$. In absence of external magnetic field, the damping of electromagnetic wave occurs, probably, due to the actions in the wave itself. In absence of magnetic field $(u_A=0)$, $\omega_{s}$ corresponds to the sound wave without damping.

\subsection{Dispersion curves for different input parameters}
	
For making numerical estimations, let us introduce following dimensionless parameters
	\begin{equation}
	x=\frac{\omega}{kc},\,\,\, \varkappa=\frac{k}{k_0},\,\,\, s=\frac{c_s}{c},\,\,\,
	a=\frac{u_A}{c},\,\,\, \nu=\frac{c}{k_0 \nu_m}.
	\label{param}
	\end{equation}
Here $k_0$ is an arbitrary scale factor for the absolute value of the wave vector $k$. By using dimensionless parameters \eqref{param}, the dispersion equation \eqref{disprel_trans} reads
	\begin{equation}
    (x^2-1)(x^2-s^2)-i\frac{\nu}{\varkappa}x[(s^2-x^2)+a^2(1-x^2)]=0.
	\label{disprel1}
	\end{equation}
Solving this equation for the function $x(\nu / \varkappa)=x_r(\nu / \varkappa)+ix_i(\nu / \varkappa)$, we obtain real and complex parts, $x_r(\nu / \varkappa)$ and $x_i(\nu / \varkappa)$, respectively. These functions define the dispersion relation curves for different modes of linear waves in the uniform (constant $\nu$) magnetized plasma for a fixed, arbitrary set of parameters $s=\frac{c_s}{c},\,\,\,a=\frac{u_A}{c},\,\,\, \nu=\frac{c}{k_0 \nu_m}$.
	
Relation \eqref{disprel1}, at varying value of $\nu/\varkappa$, can be considered as equation, which describes propagation of  the wave at a fixed $\varkappa$ through the medium with a variable $\nu$.
	
In the medium with smoothly changing parameters the short wave propagation is approximately described by Eq.\eqref{disprel1}. It may be interpreted as WKB approximation \cite{wkb} describing propagation of waves with a fixed $k$, which length $\lambda$ is much less than the characteristic length of varying of parameters, $\lambda \ll \lambda_\rho =\rho/|\nabla \rho|$ and $\lambda \ll \lambda_H =  H/|\nabla H|$, in a non-uniform medium. Interpretation of wave transformation in the non-uniform media made below needs more detailed and careful analysis, than the one, which is based on the simplified, approximate equation \eqref{disprel1}.
	
Depending on parameters $a$ and $s$, different wavemodes correspond to various waves, and conversion of waves happens when the wave is propagating through the region with varying $\nu$.
We deal here with 4 types of waves:
\begin{enumerate}
	\item Sound waves (SW).
	
	\item Electromagnetic waves (EM).
	
	\item Fast MHD waves.
	
	\item Slow MHD waves,
	
\end{enumerate}
which in the case of a uniform  magnetic field perpendicular to the direction of the wave propagation  are reduced to static perturbation \cite{kulik,lleldin}.
 
The waves here differ from classical EM waves in vacuum, or waves in the ideal MHD, which do not have imaginary parts. Here EM wave is damping in the media without magnetic field (\ref{decremem}), and the speed of the fast MHD wave (\ref{fast_slow}) differs from its classical value \cite{kulik,lleldin} due to account of the displacement current.
	
Usually in the experiments and devices on the Earth the relation between non-dimensional parameters (\ref{param}) is $a\ll 1$, $s\ll 1$, but their ratio may be arbitrary. In astrophysical objects, like pulsars, X-ray sources, AGNs   relativistic effects could be important and we may expect $a, s \lesssim 1$. 
 
\begin{figure}
		\centering
		\includegraphics[width=13.0cm,height=8.1cm]{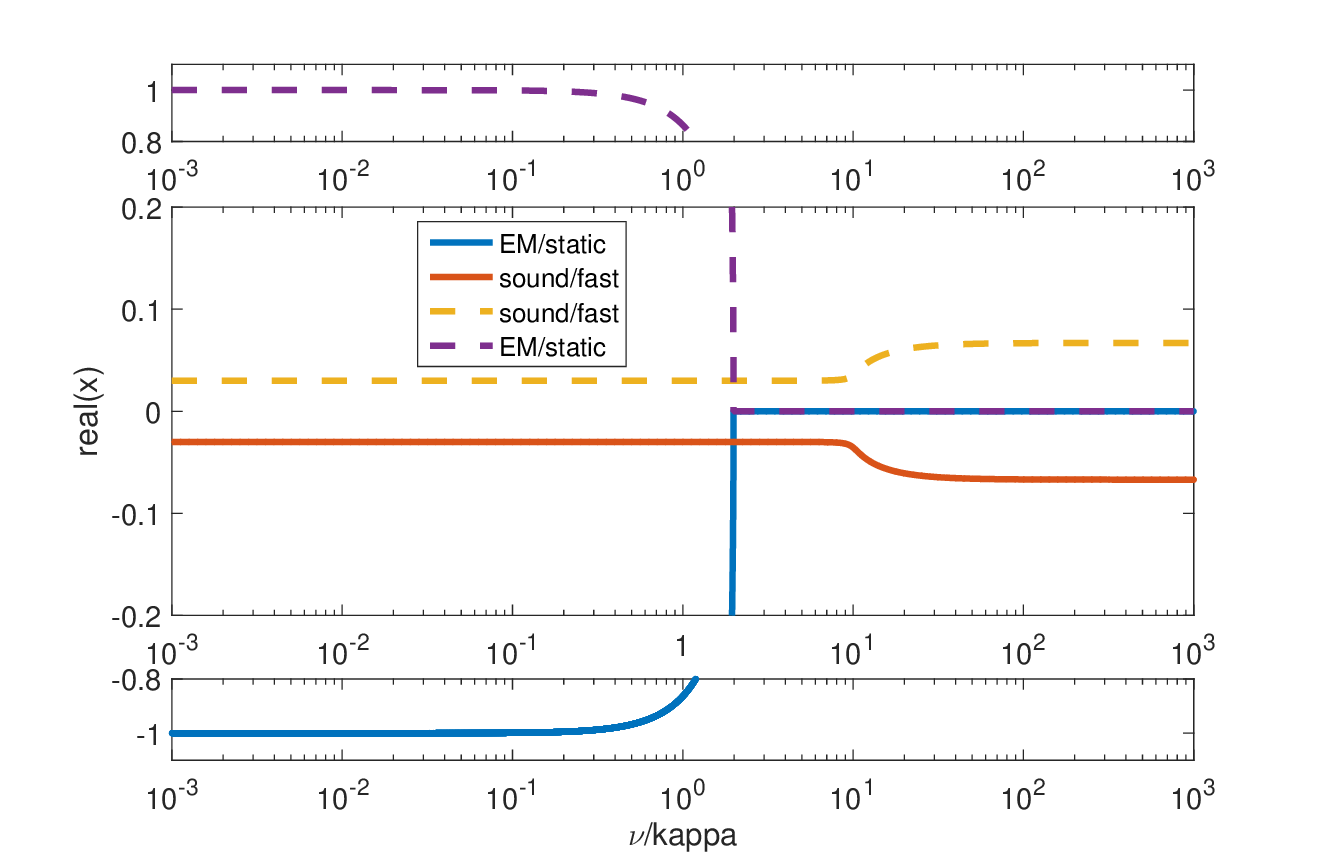}
		\caption{ Real parts of solutions of eq.\eqref{disprel1} for parameters $s = 0.03,  a = 0.06$. Blue and purple curves correspond to electromagnetic waves, and  static perturbation (slow MHD). Yellow and red ones correspond to sound and fast MHD waves. Considering waves with a fixed $\varkappa$, we describe qualitatively transformation of types of waves, propagating in medium with varying $\nu$, see discussion in the text. The region within $|x_r| \in [0.2,0.8]$, where only damping EM waves are presented, is excluded from the plot to improve its readability.
  }
		\label{fig:case1r}
	\end{figure}

\begin{figure}
		\centering
		\includegraphics[width=13.0cm,height=8.1cm]{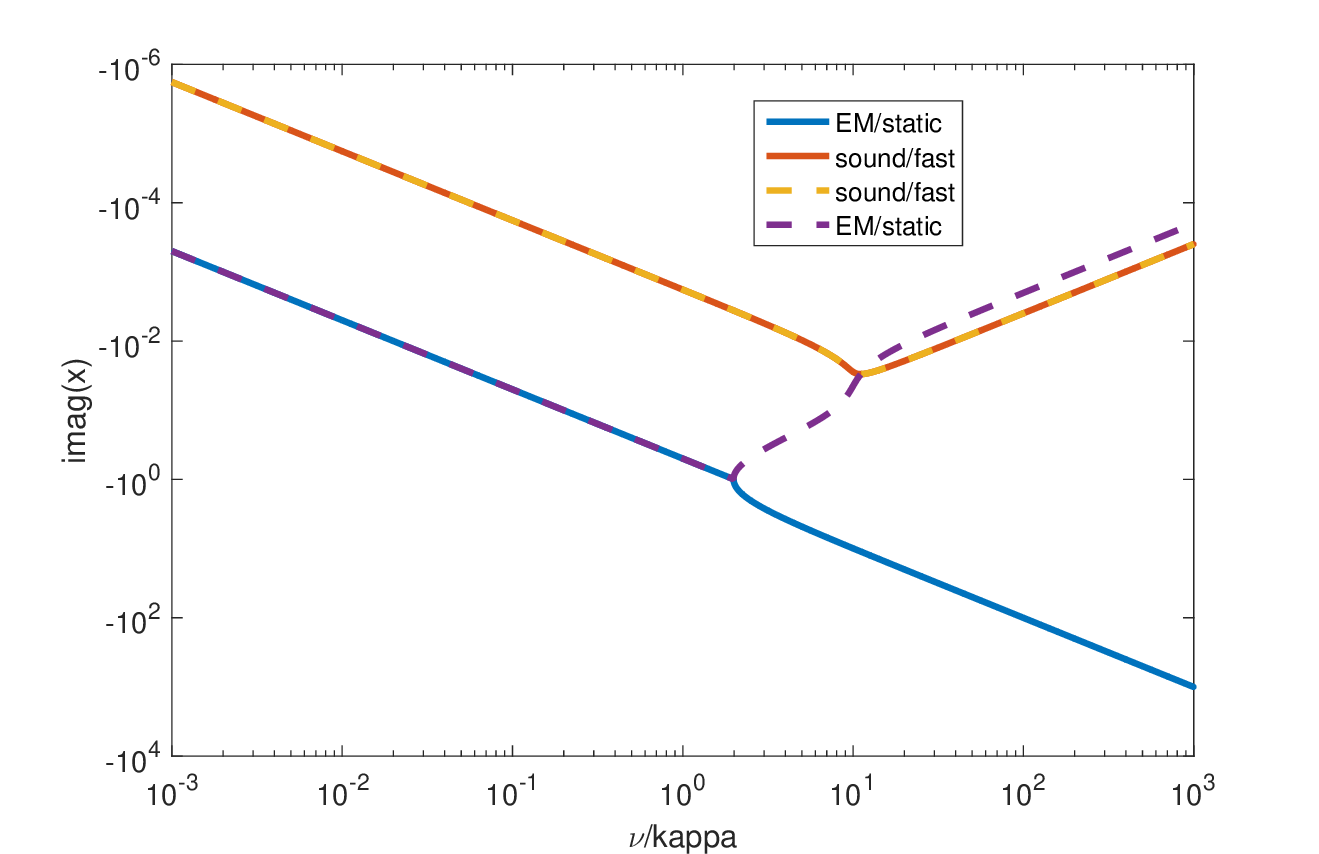}
		\caption{Imaginary parts of solutions of eq.\eqref{disprel1} for parameters $s = 0.03, a = 0.06$.
			Blue and purple curves correspond to electromagnetic waves and  static perturbation (slow MHD). Yellow and red ones correspond to sound and fast MHD waves.  
		}
		\label{fig:case1i}
	\end{figure}

For better qualitative understanding of the behaviour of different modes from the dispersion equation (\ref{disprel1}) we start from  the the solution with moderate input parameters $s = 0.03$, $a = 0.06$, and find for these parameters  numerical solutions $x_r(\nu/\varkappa),\,\,x_i(\nu/\varkappa)$, for four different modes presented in Figs.\ref{fig:case1r},\ref{fig:case1i}.

In this case, rapid MHD waves are smoothly connected with the sound ones.  Additional calculations had shown, that the electromagnetic waves exist at $\nu/\varkappa = 2$ for any $a,s\ll 1$, and don't propagate in the region with $\nu/\varkappa > 2$. 
	
Similar topology of wave modes takes place for $a<a_{b} \simeq 2.82s$ for low $s\ll 1$, this ratio changes with increase of $s$ to $a_{b}\simeq 2.75s$ at $s = 0.1$. For increasing values of $a$ and $s$, the qualitative presentation similar to Figs.\ref{fig:case1r},\ref{fig:case1i} takes place for Alfven velocities below the threshold value $ a_{thr}$ of normalized Alfven velocity, depending on $s$ (see Table \ref{tab:table1_rel}). 
		
	\begin{figure}
		\centering
		\includegraphics[width=13.0cm,height=8.1cm]{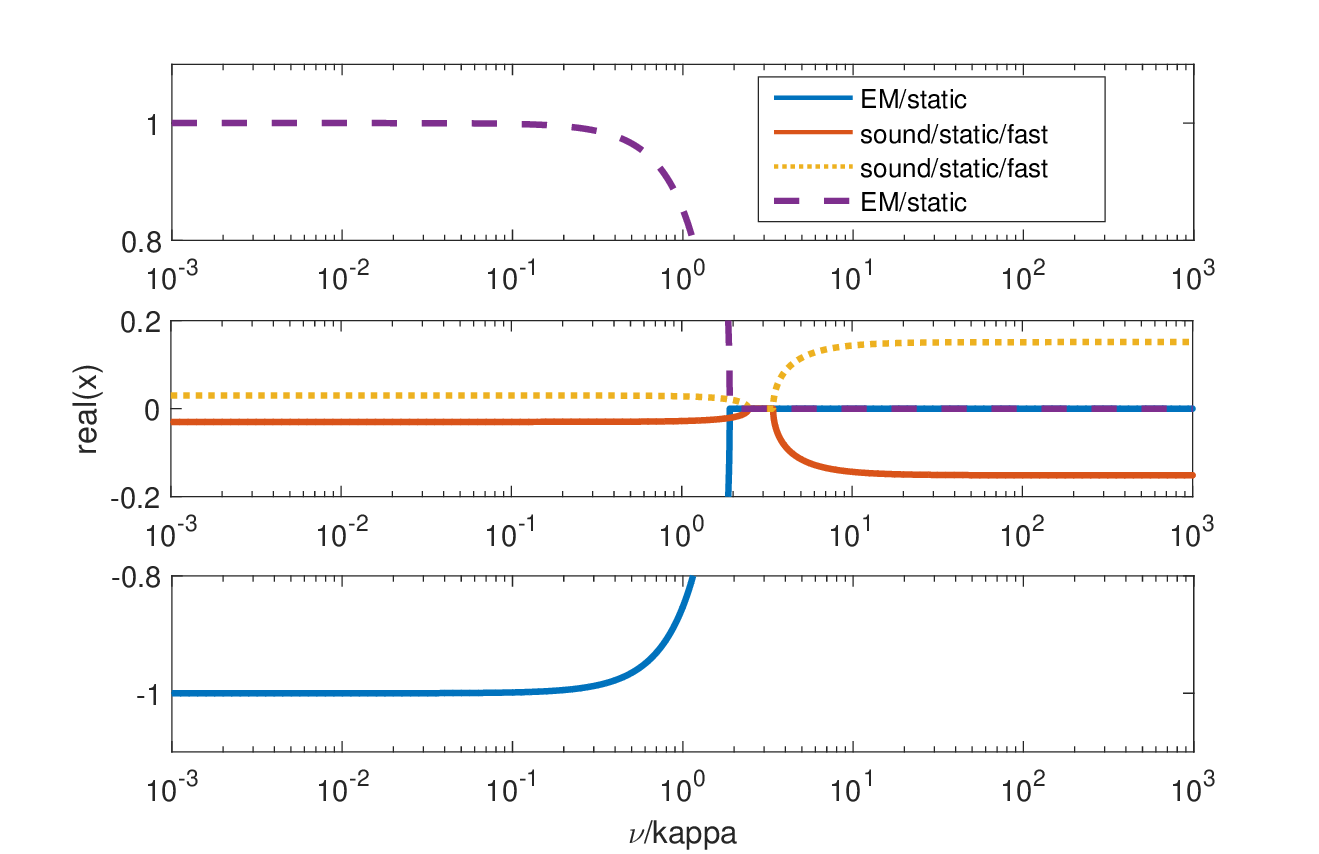}
		\caption{
			Real parts of solutions of eq.\eqref{disprel1}
			for parameters $s = 0.03$, $a = 0.15$. Blue and purple curves correspond to electromagnetic  waves and static perturbation. Yellow and red ones correspond to the sound and fast MHD waves. At used here input parameters there is an interval of the values of  $\nu/\varkappa$, where only damping static perturbations may exist. The region within $|x_r| \in [0.2,0.8]$, where only damping EM waves are presented, is excluded from the plot to improve its readability.
		}	
		\label{fig:case2r}
	\end{figure}
	
	\begin{figure}
		\centering
		\includegraphics[width=13.0cm,height=8.1cm]{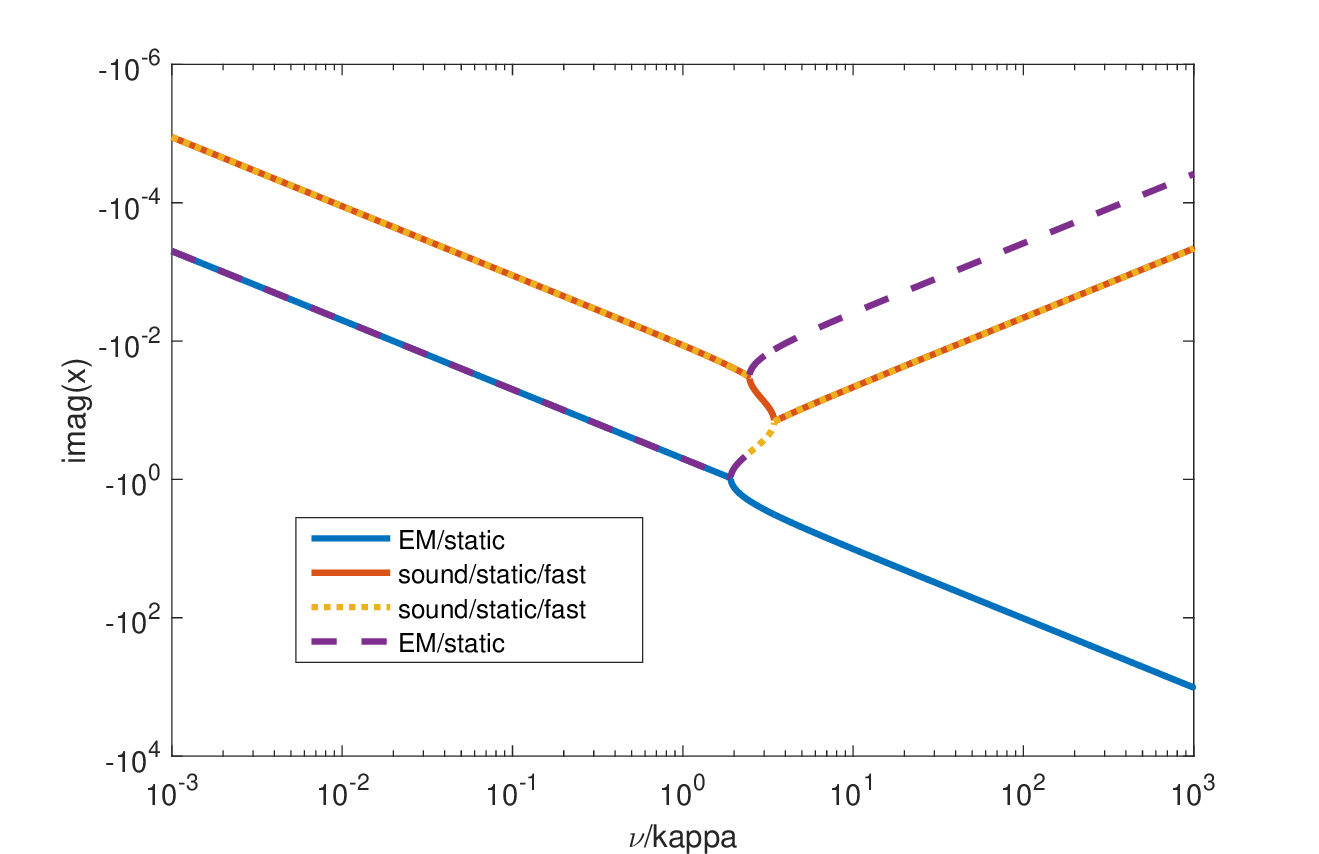}
		\caption{Imaginary parts of solutions of eq.\eqref{disprel1} for parameters $s = 0.03$, $a = 0.15$. Blue and purple curves correspond to electromagnetic wave and  static perturbation. Yellow and red ones correspond to the sound and fast MHD waves. 			
		}	\label{fig:case2i}
	\end{figure}
 
The second case differs from the first one due to appearance of a region, where all four wavemodes have zero real parts $x_r = 0$. It is the region, where all waves are damped. In the Figs.\ref{fig:case2r},\ref{fig:case2i} the real and imaginary parts of $x$ for the second case are plotted for $s = 0.03$, $a = 0.15$.

\begin{figure}
		\centering
		\includegraphics[width=13.0cm,height=8.1cm]{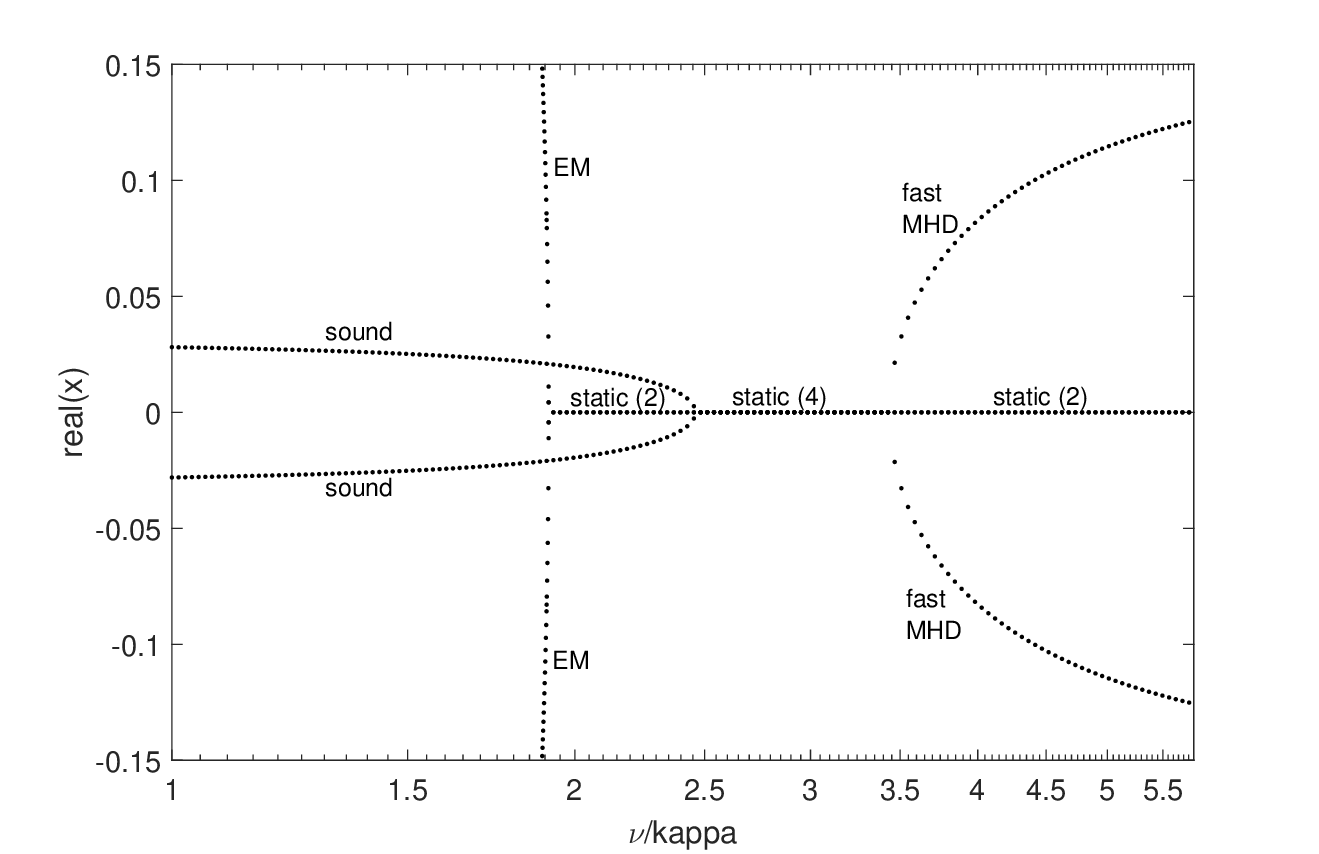}
		\caption{
			The vicinity of wave transitions region (Case 2, real parts) for parameters $s = 0.03$, $a = 0.15$ from the Fig.\ref{fig:case2r}. The region near the transition point of EM waves has fine grid to resolve large gradients in the solution. On the left side of the plot, the damping EM and sound waves are shown. On the right side there are MHD waves and  static perturbation. Both regimes are connected with the region, where all waves are damped.
		 }	
		\label{fig:dotsr}
	\end{figure}

\begin{figure}
		\centering
		\includegraphics[width=13.0cm,height=8.1cm]{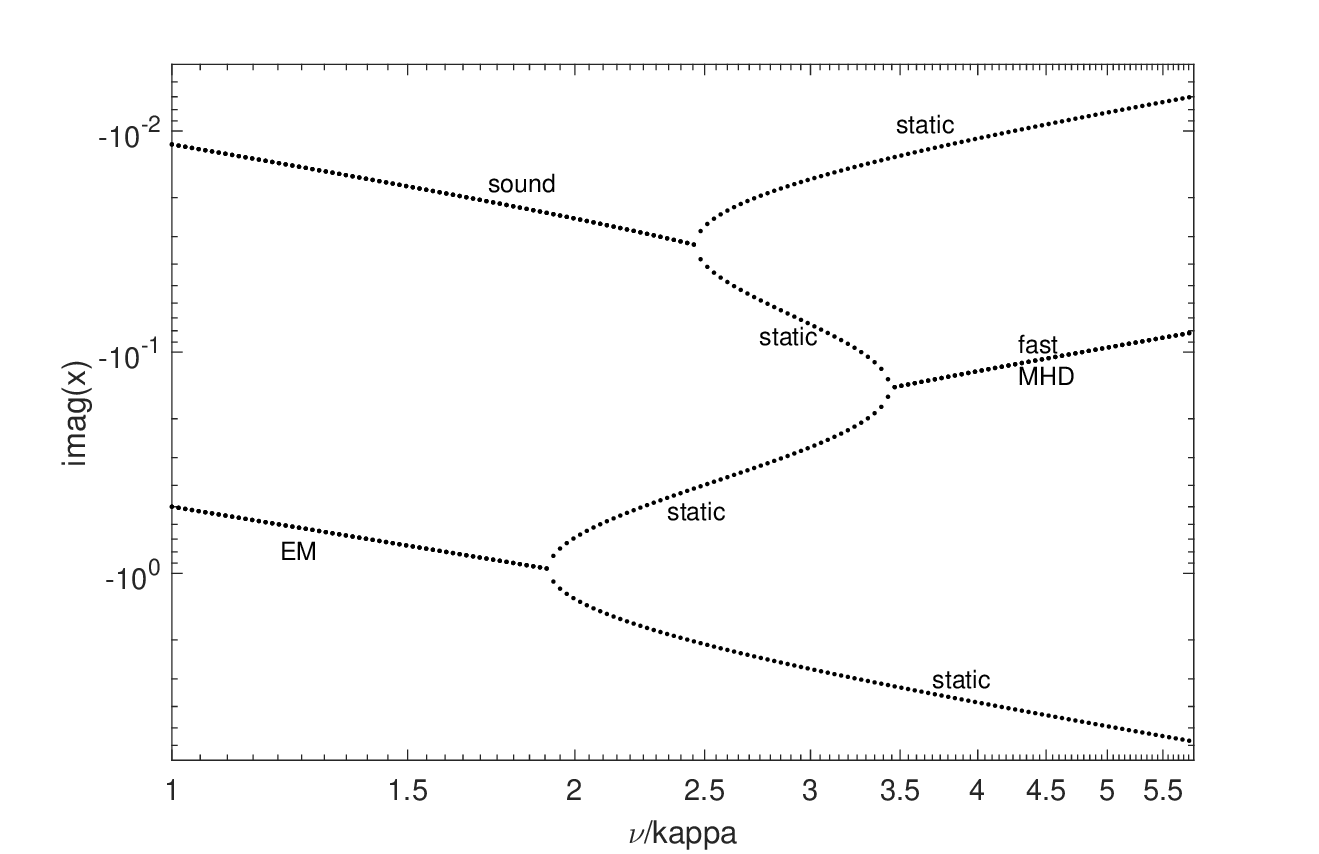}
		\caption{
			The vicinity of wave transitions region (Case 2, imaginary parts) for parameters $s = 0.03$, $a = 0.15$ from the Fig.\ref{fig:case2i}. On the left side of the plot, the acoustic (upper) and EM (bottom) decrements are shown. On the right side two  decrements (upper and bottom) for  static perturbations and one for fast MHD waves (middle) are plotted. These two plots are connected with the region without waves.
		}	
		\label{fig:dotsi}
	\end{figure}

The existence of the region without waves takes place, if $a > a_{b}\simeq 2.82s$ at $a\ll 1$ for low $s$ values, i.e. in highly magnetized plasma. With increasing of $s$, the value of $a_{b}$ changes up to the value $\simeq 2.75 s$.  At small  $\nu/\varkappa$ we have sound  and electromagnetic waves. It this case the structure of solutions of the dispersion equation \eqref{disprel1} is complicated in the vicinity of the transition region at $\nu/\varkappa \sim 3$. It is presented at larger scale  in Figs. \ref{fig:dotsr}, \ref{fig:dotsi} by points with four solutions corresponding to each value of $\nu/\varkappa$.

 \begin{table}
		\begin{center}
			\caption{
   Threshold values of normalized Alfven speed for given sound speeds. At larger values of a the electromagnetic wave is forming directly from the fast MHD wave (Case 3). 
    It may be seen, that for $s$ values of lower than $0.1$, $a_{thr}$ is almost constant,  and for larger values, the ratio $a_{thr}/s$ decreases. }
			\label{tab:table1_rel}
			\begin{tabular}{l|r}
				\textbf{$s = c_s/c$} & \textbf{$a_{thr} = u_A/c$}\\
				\hline
				0.01 & 0.35 \\
				0.10 & 0.36 \\
				0.20 & 0.45 \\
				0.30 & 0.54 \\
				0.40 & 0.62 \\
				0.50 & 0.70 \\	
				0.60 & 0.77 \\			
			\end{tabular}
		\end{center}
	\end{table}

The third case corresponds to larger values of $a$. The solutions of Eq.\eqref{disprel1} starts to deviate from the first two cases at $a>0.35$, where the electromagnetic wave is directly connected with the fast MHD wave at $a>s$.  Such conditions may appear during accretion processes onto highly magnetized neutron star and during accretion of the magnetized matter into a black hole in galactic sources and AGN's \cite{bkuniv}.

	\begin{figure}
		\centering
		\includegraphics[width=13.0cm,height=8.1cm]{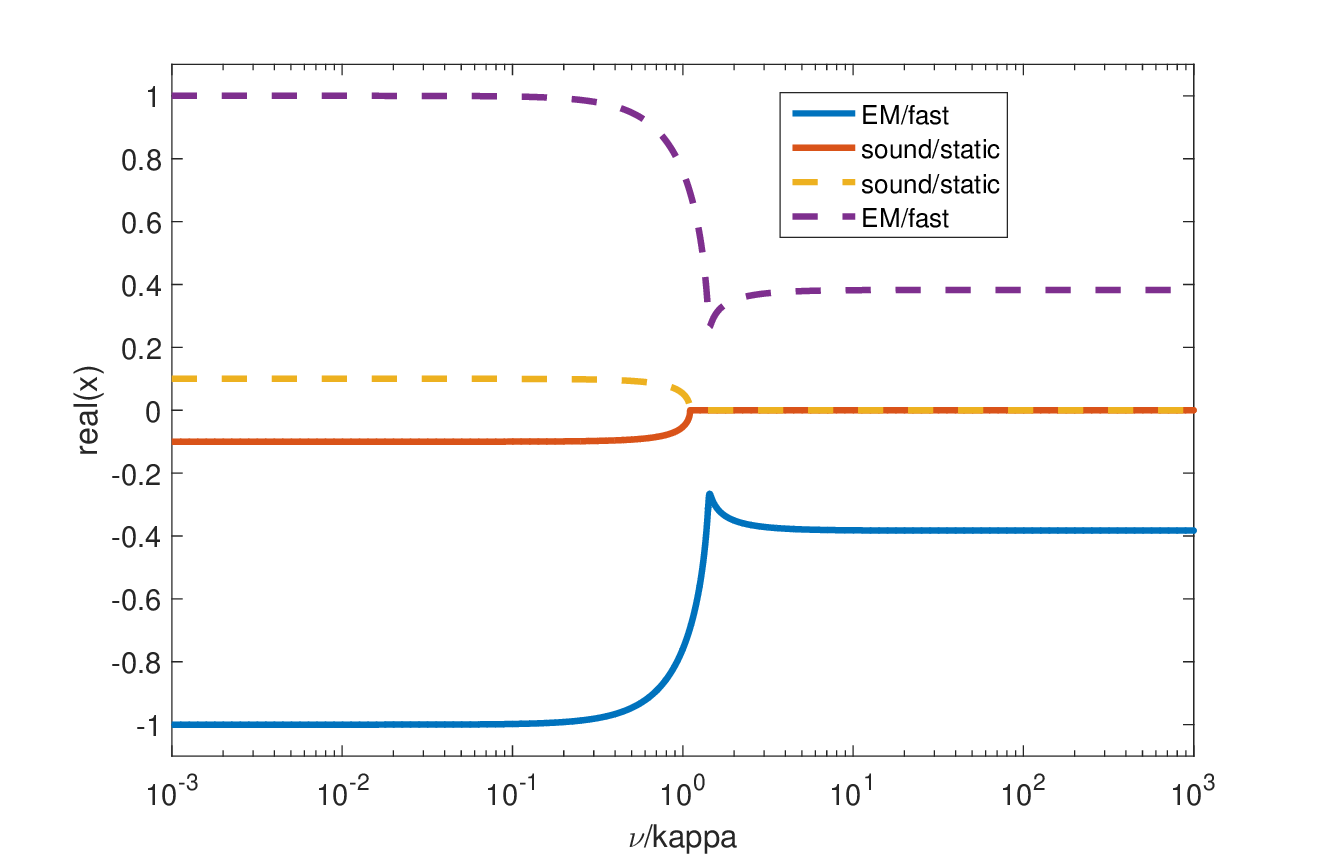}
		\caption{Real parts of solutions of eq.\eqref{disprel1}
			for parameters $s = 0.1, a = 0.4$. Blue and purple curves correspond to electromagnetic and fast MHD waves. Yellow and red ones correspond to sound waves and  static perturbations.
		}
		\label{fig:case3r}
	\end{figure}

\begin{figure}
		\centering
		\includegraphics[width=13.0cm,height=8.1cm]{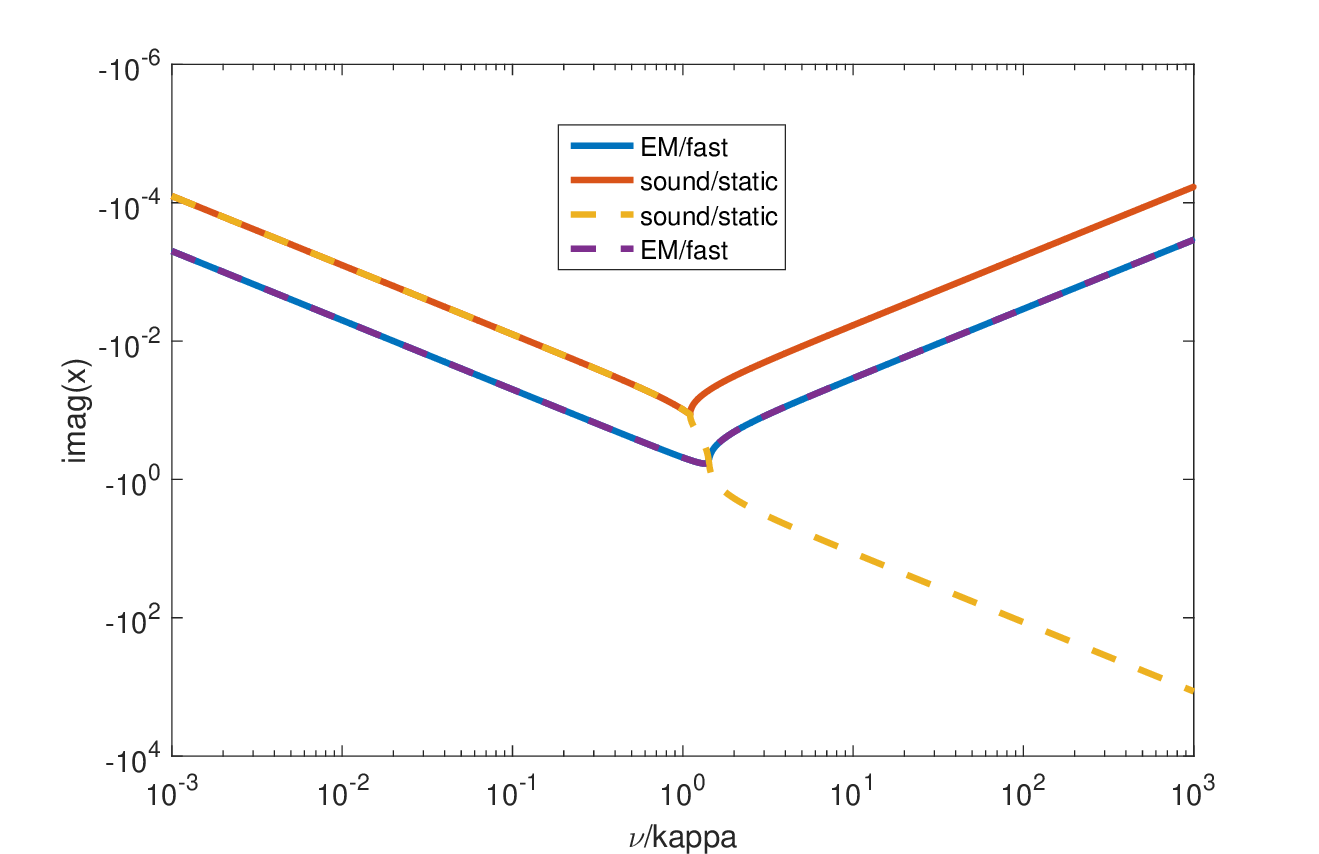}
		\caption{Imaginary parts of solutions of eq.\eqref{disprel1} for parameters $s = 0.1, a = 0.4$. Blue and purple curves correspond to electromagnetic and fast MHD waves. Yellow and red ones correspond to sound and  static perturbations.	
		}
		\label{fig:case3i}
	\end{figure}

  The sound waves don't penetrate into region with $\nu/\varkappa\geq\sim 1$. In  Figs.\ref{fig:case3r},\ref{fig:case3i} this regime takes place at $s = 0.1$, $a = 0.4$. The connection of electromagnetic waves with fast MHD wave occurs at $1 < \nu/\varkappa < 2$. Table \ref{tab:table1_rel} shows boundary values of Alfven speed $a_{thr}$ as a function of the sound speed $s$. At $a>a_{thr}$ we have the Case 3,  and at $a<a_{thr}$ Cases 1 or 2 take place.

\section{Waves propagation along the magnetic field}

\begin{figure}
		\centering
		\includegraphics[width=13.0cm,height=8.1cm]{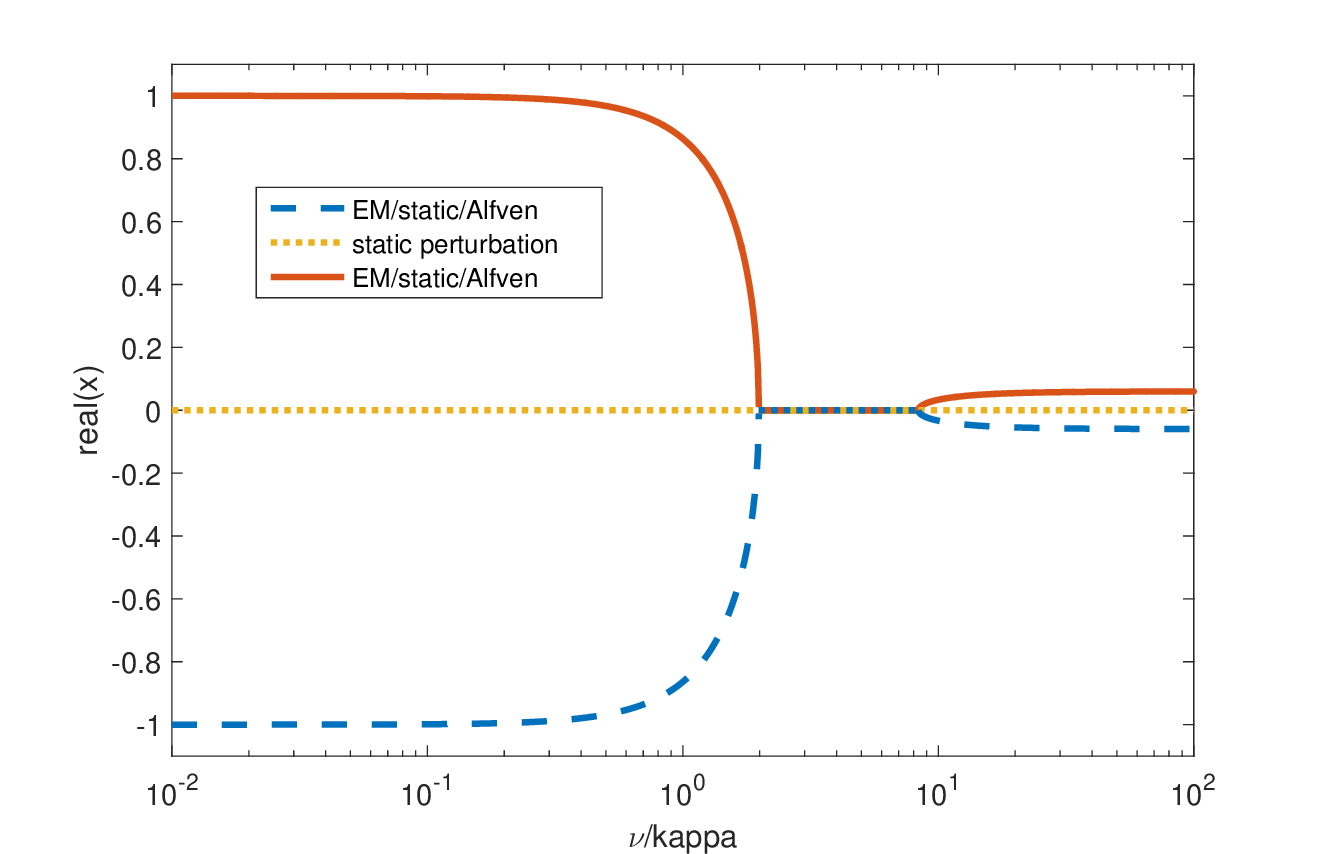}
		\caption{Real parts of solutions of Eq.\eqref{disprel_par} (without sound waves) for $a = 0.06$. Blue and red curves on the left side of the plot correspond to EM waves, the yellow one corresponds to a static perturbation. On the right side of the picture, red and blue curves correspond to Alfven waves, the yellow line corresponds to static perturbation. The transformation of waves does not happen because of existence of the intermediate region with damping static perturbations. }
		\label{fig:alf1r}
	\end{figure}

\begin{figure}
		\centering
		\includegraphics[width=13.0cm,height=8.1cm]{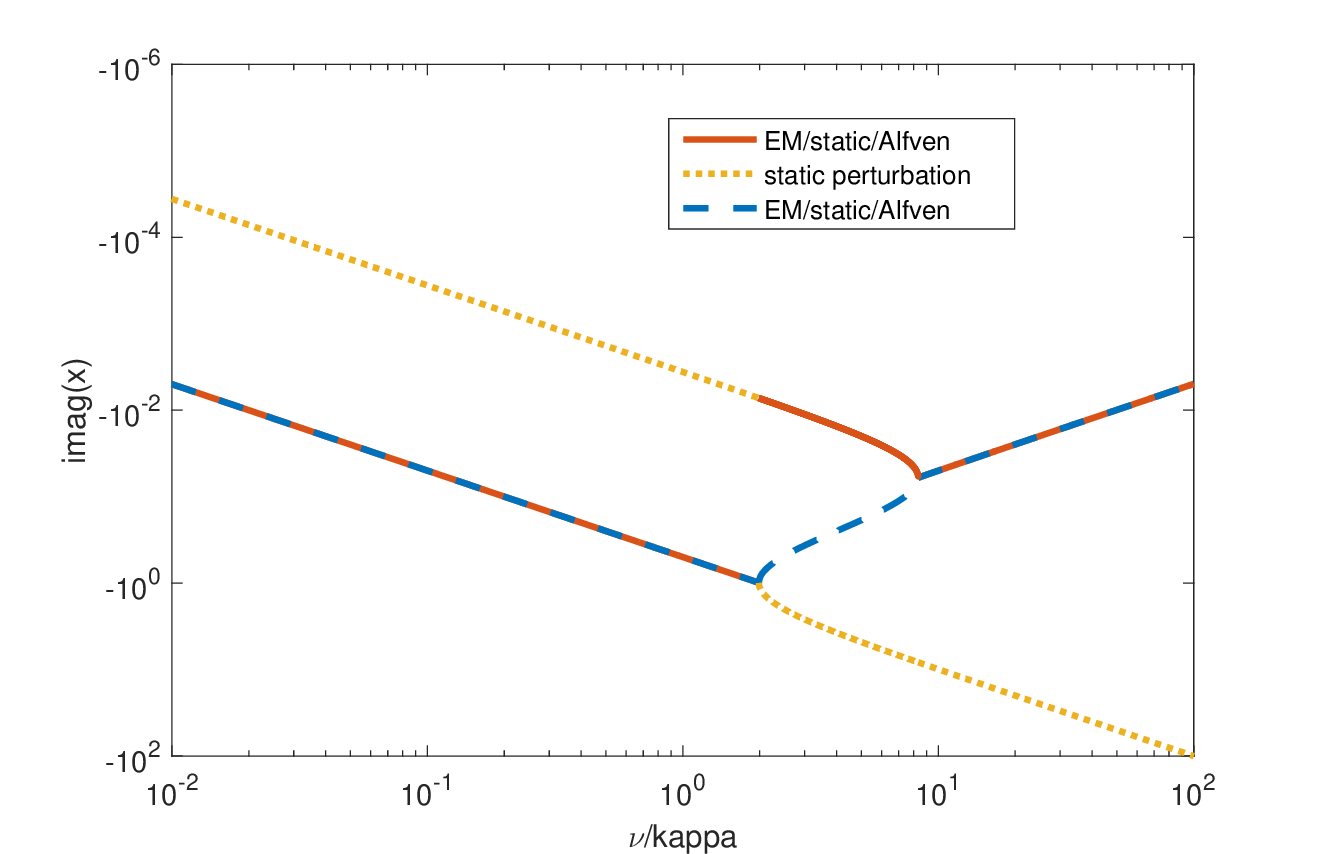}
		\caption{Imaginary parts of solutions of Eq.\eqref{disprel_par} (without sound waves)  for $a = 0.06$, see caption to the Fig.\ref{fig:alf1r}.}
		\label{fig:alf1i}
	\end{figure}

\begin{figure}
		\centering
		\includegraphics[width=13.0cm,height=8.1cm]{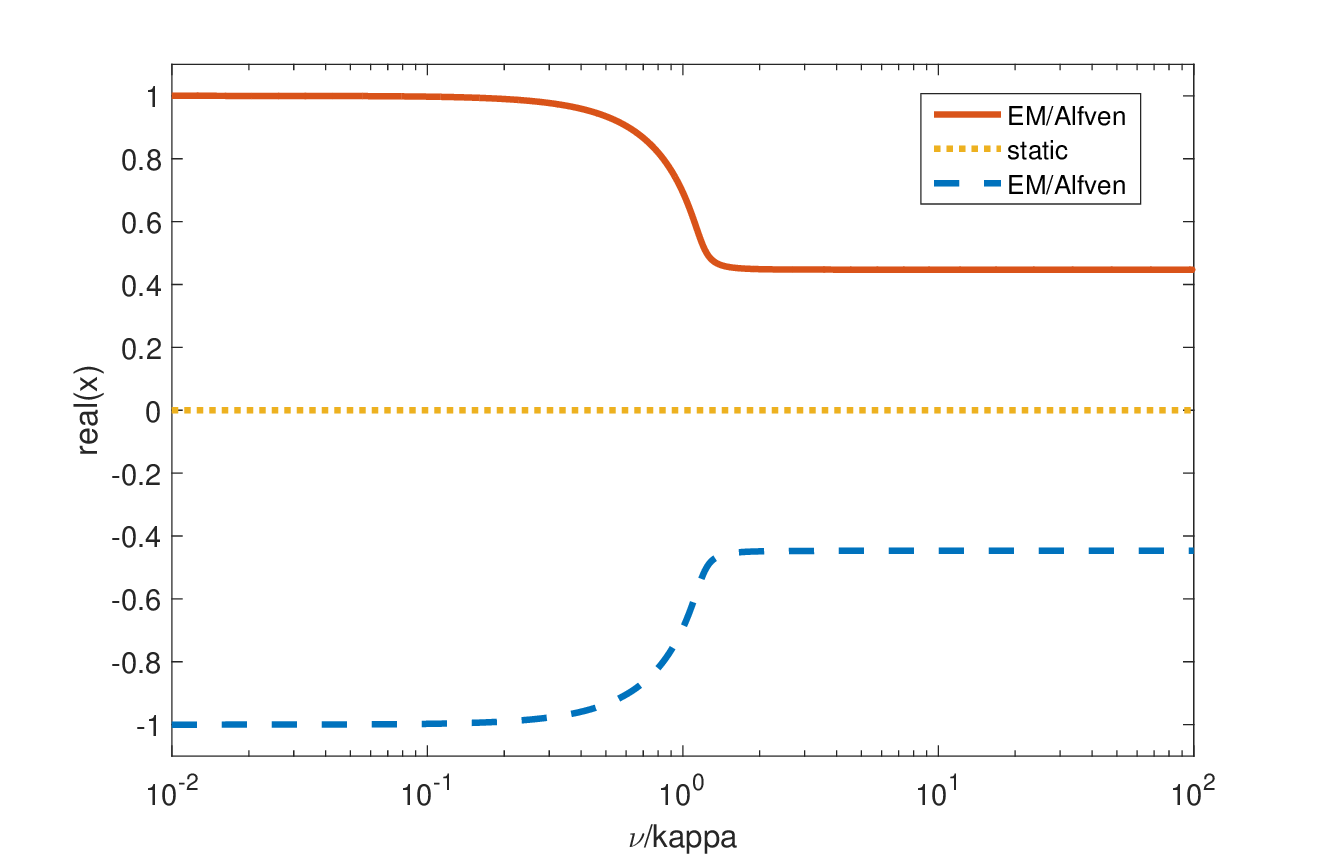}
		\caption{Real parts of solutions of eq.\eqref{disprel_par} (without sound waves) for $a = 0.5$. Blue and red curves correspond to EM and Alfven waves. Yellow line corresponds to static perturbations.}
		\label{fig:alf2r}
	\end{figure}

\begin{figure}
		\centering
		\includegraphics[width=13.0cm,height=8.1cm]{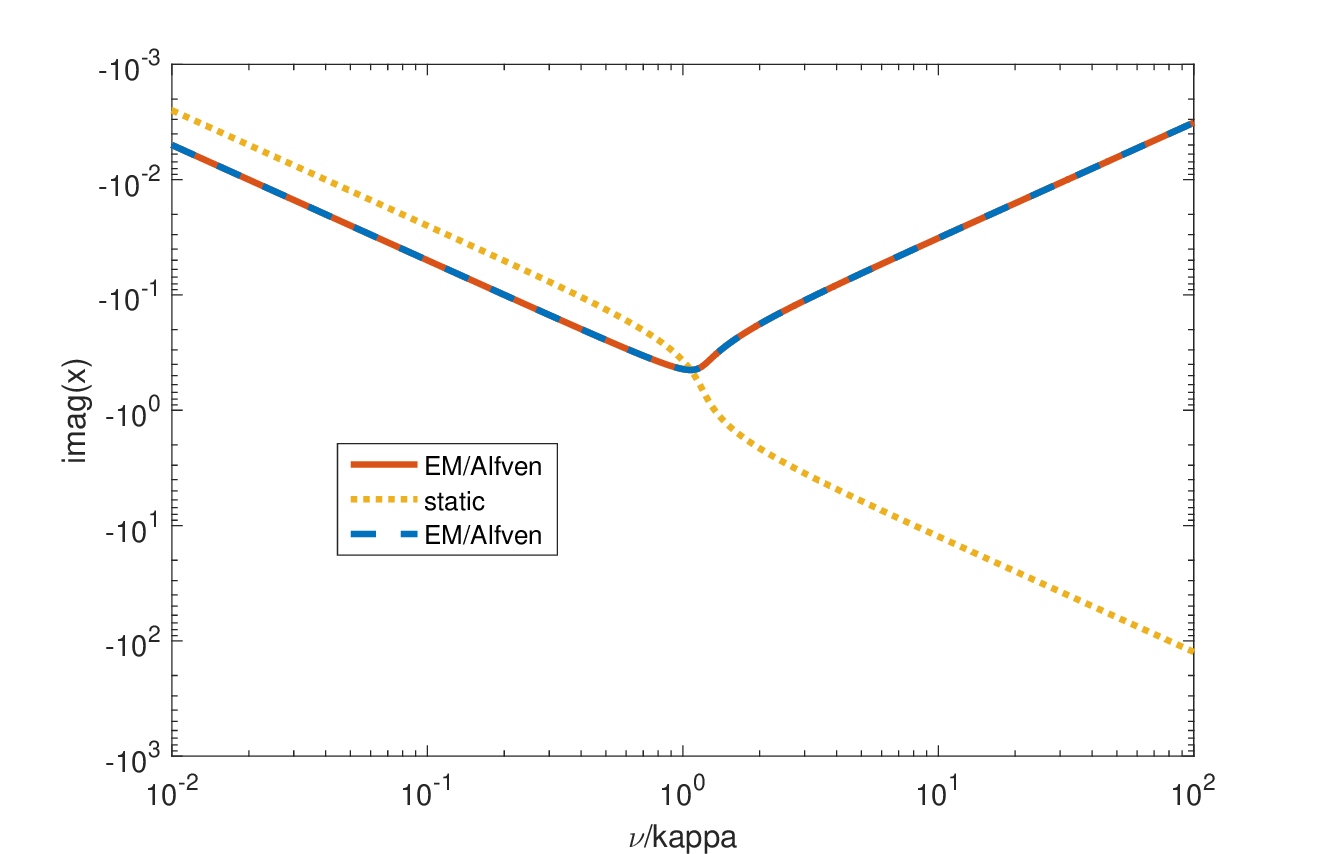}
		\caption{Imaginary parts of solutions of eq.\eqref{disprel_par} (without sound waves) for $a = 0.5$. Blue and red curves correspond to EM and Alfven waves. Yellow line corresponds to static perturbations.}
		\label{fig:alf2i}
	\end{figure}

Waves, propagating in the direction along the magnetic field have $k_y=0,\,\, k_x^2=k^2$, and from \eqref{disprel} we obtain the dispersion equation for these waves in the form 

\begin{eqnarray}
	(k^2c^2-\omega^2)(k^2c_s^2-\omega^2)-\nonumber \\
 i\frac{\omega c^2}{\nu_m}\left[k^2(c_s^2+u_A^2)-\omega^2\left(1+\frac{u_A^2}{c^2}\right)+k^2c_s^2\frac{u_A^2}{c^2}\left(1-\frac{k^2c^2}{\omega^2}\right)\right]=0. 
 \label{disprela}
	\end{eqnarray}
After some algebraic transformations this equation is written in the form
    
\begin{equation}
    (k^2c_s^2-\omega^2) \left\{ k^2c^2-\omega^2 - i\frac{c^2}{\omega \nu_m} \left[ \omega^2 \left(1+\frac{u_A^2}{c^2}\right)- k^2 u_A^2 \right] \right\} = 0,
    \label{disprel_parallel}
\end{equation}
For this configuration the sound  waves don't depend on waves, connected with magnetic field. The last ones consist of the electromagnetic waves propagating with the light speed $|\omega_e|=k_x c$ for dielectric conditions, with very large magnetic viscosity $\nu_m >>c^2/\omega$; and Alfven wave, propagating with Alfven speed, where energy of the wave is taken into account, so that $\omega=\frac{ k_x u_A}{\sqrt{1+u_A^2/c^2}}$, for ideal plasma with $\nu_m=0$. The mixture of these two waves for intermediate $\nu_m$ is shown in the Figs.\ref{fig:alf1r}-\ref{fig:alf2i}.  


In the case of almost ideal plasma $(\nu_m \ll c^2/ \omega)$, one can write the dispersion equation \eqref{disprel_parallel} in the form

 \begin{equation}
    (k^2c_s^2-\omega^2) \left[ \omega^2 \left(1+\frac{u_A^2}{c^2}\right)- k^2 u_A^2 \right] = 0.
    \label{mhdparal}
\end{equation}  

Four roots of the ideal MHD equation \eqref{mhdparal} are written as 

    \begin{equation}
\omega_{A0}^2 = \frac{k^2u_A^2}{1+u_A^2/c^2},  \quad \omega_s^2=k^2 c_s^2.
	\label{mhdroots}
	\end{equation}
In absence of matter viscosity, in the linear approximation the sound waves  propagate along the  magnetic field without dissipation.   

Alfven waves damping at non-zero $\nu_m$ is found from \eqref{disprel_parallel} as

    \begin{equation}
	\omega_A= \omega_{A0} + i\gamma_A,  \quad  \gamma_A = -\nu_m \frac{k^2 c^4}{2(c^2+u_A^2)^2}
	\label{decrem_alfven}
	\end{equation}
Here we have also a rapidly damping static perturbations, similar to \eqref{slowdamp_2}.

In the case of plasma with very low conductivity $(\nu_m \gg  c^2/ \omega )$ there is a solution of the equation \eqref{disprel_parallel} in the form of slowly damping  electromagnetic waves (EM), and static perturbations, assuming that their decrements are smaller than $ku_A, kc$:
	\begin{equation}
		\omega_{em} = kc - i\frac{c^2}{2\nu_m}, \qquad \omega_{S*} = 0 - i\frac{u_A^2}{\nu_m}.	
	\label{decrem_along}
	\end{equation}
The electromagnetic wave propagation, and behaviour of the static perturbations don't depend on the magnetic field direction, so the increments $\gamma_{em}$ from \eqref{decrem_along}  and \eqref{decremem} coincide.


Depending on the parameter $a$, different modes along the field correspond to various waves, a conversion of the waves may happen when the they propagate through the region with varying $\nu$. We deal here with 4 types of waves:

\begin{enumerate}
	\item Non-damping sound waves.
	
	\item Electromagnetic waves with damping.
	
	\item Alfven waves.
	
	\item Static perturbations. 
\end{enumerate}
 
Here we don't consider sound waves, because their properties are the same for all values of $a, \nu_m$. We use here the same WKB-type analysis for estimation of wave conversions.

In non-dimensional units \eqref{param}, we obtain from \eqref{disprel_parallel} a dispersion equation in the form

\begin{equation}
	x(x^2-1)-i\frac{\nu}{\varkappa}(1+a^2)\left(\frac{a^2}{1+a^2}-x^2\right) = 0.
	\label{disprel_par}
\end{equation}
In  Figs.\ref{fig:alf1r},\ref{fig:alf1i} the solution of equation \eqref{disprel_par} is plotted for $a = 0.06$. Like in the field-transverse case, a region is presented, where the Alfven and EM waves do not exist, and only static perturbations, and sound waves are present there.

Calculations show that for $a \gtrsim 0.35$ EM waves are forming directly from the Alfven ones, see Figs. \ref{fig:alf2r}, \ref{fig:alf2i} for $a = 0.5$. 
 
\section{Dispersion curves for the electromagnetic branch in Eq. \eqref{disprel_1}}

The equation \eqref{disprel_1} is not linear independent, because it follows from \eqref{disprel}  at $u_A=0$. It describes  electromagnetic waves in a non-magnetized plasma.  This equation is solved exactly, giving

\begin{equation}
    \omega = \pm kc\sqrt{1 - \frac{c^2}{4k^2\nu_m^2}} - i\frac{c^2}{2\nu_m}, 
\end{equation}
The electromagnetic wave with small damping is propagating in a media with low conductivity (dielectric, at $\nu_m\gg \frac{c}{2k}=\frac{c^2}{2\omega}$), repeating the limiting case in Eqs.\eqref{additionem},\eqref{decremem}. 

\begin{figure}
		\centering
		\includegraphics[width=13.0cm,height=8.1cm]{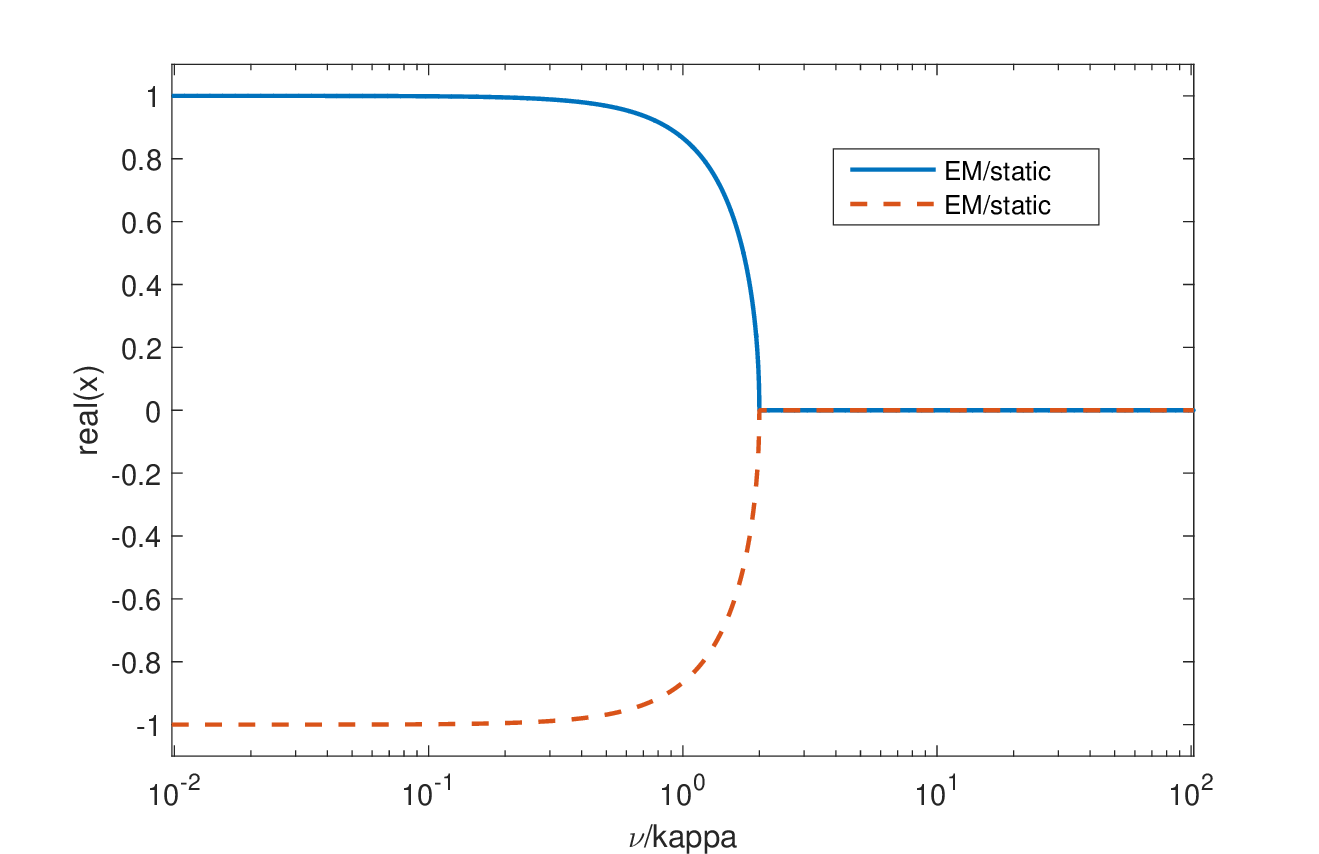}
		\caption{Real parts of solutions of Eq. \eqref{disprel_1}. Blue and red curves mark different roots, corresponding to EM waves with damping for $\nu/\varkappa < 2$, and static perturbations for larger values. }
		\label{fig:em_real}
	\end{figure}

\begin{figure}
		\centering
		\includegraphics[width=13.0cm,height=8.1cm]{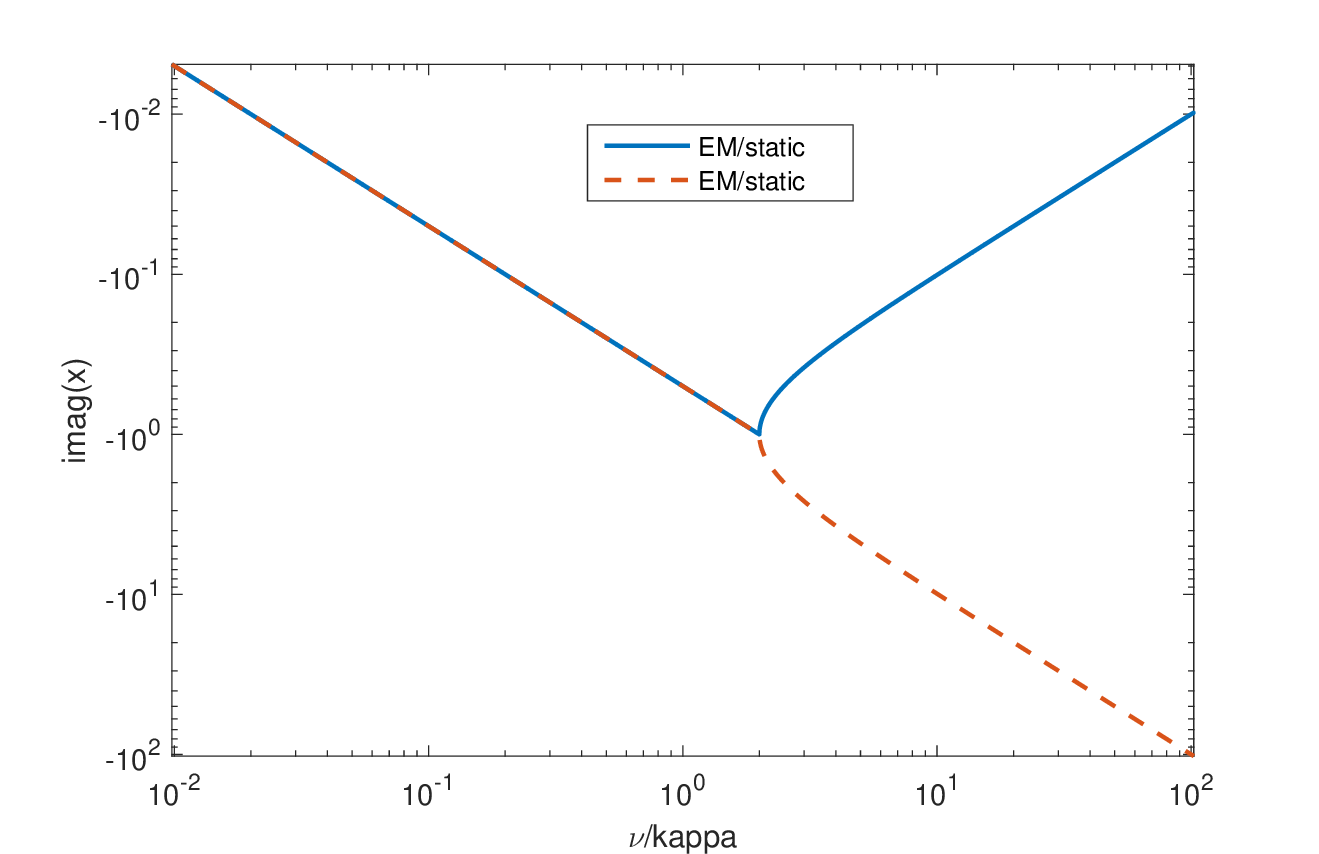}
		\caption{Imaginary parts of solutions of Eq. \eqref{disprel_1}. Blue and red curves mark different roots, corresponding to EM waves with damping for $\nu/\varkappa<2$ and static perturbations for larger values. }
		\label{fig:em_imag}
	\end{figure}

In non-dimensional units \eqref{param}, this solution is plotted as a function of $\nu/\varkappa$ in the Figs \ref{fig:em_real} (real parts) and \ref{fig:em_imag} (imaginary parts). Appearance of the static perturbation instead of the electromagnetic wave occurs at $\frac{c}{2k\nu_m} = 2$, similar to Fig.\ref{fig:case1r}. Static perturbations in the highly conducting plasma have modes of slow and rapid damping. 

\section{Discussion and Conclusions}

The MHD equations with vortex electrical field (MHDE) are solved here for a problem of propagation of linear waves in the magnetized uniform plasma at finite scalar constant conductivity.

We have derived a dispersion equation for different  linear modes in this condition. Solutions of this equation had been obtained analytically for limiting cases of very high and very low electrical conductivity. Numerical solutions had been obtained for intermediate values of this parameter.

Solutions of this equation include different physical  modes, namely, hydrodynamic (sound) waves, MHD waves, electromagnetic waves, damping static perturbations, and waves with intermediate properties. Both real and imaginary parts of these modes have been found, and presented in figures.  

Parameters, characterizing the wave vector $\varkappa$, and plasma conductivity $\nu$ are present in the dispersion equation in the form of their ratio $\nu/\varkappa$.
Applying WKB approximation we have approximately interpreted the solutions of this equation for describing the change of a speed for different modes with constant $\varkappa$, propagating in the medium with smoothly changing conductivity.

The conclusions about the wave transformation in the medium with changing conductivity is clearly visible in many cases. In Figs. \ref{fig:case2r}, \ref{fig:alf1r}, where there are  regions with only static perturbations, the wave transformation does not occur. The evident cases are connected with transformation of rapid MHD wave into the electromagnetic one, perpendicular to the uniform magnetic field, and Alfven into electromagnetic, along the magnetic field, at propagation in the direction of decreasing conductivity parameter. We have shown, that transformation  of MHD into EM wave may happen only in a highly magnetized plasma with  $a > a_{thr}$.

We used here the classic MHD description of the continuous media like it was done also in FMHD by \cite{mcgre}. The matter involved into the consideration cold be plasma, metal, possibly liquid,  for a highly conducting media. Very resistive materials could be cold neutral gas, transparent dielectric, like glass or plastic. Very cold and relatively dense gas clouds, $"$dark$"$ and $"$black$"$ ones are observed in the Galaxy, see e.g.  \cite{rr,cloud,gwl}. These clouds consist of the neutral gas, and contain large scale magnetic fields. New stars are continuously produced in these clouds due to action of the gravitational instability. Collision of a rapid plasma wind from pulsars or blue supergiant stars with such cold magnetized cloud should lead to formation a very long electromagnetic wave, propagating through the cloud. To describe this phenomena the full non-linear MHDE set of equations should be used.

The MHDE system of equations, can be applied to mildly relativistic magnetized jets from magnetically-driven core-collapse supernovae \cite{takiwaki2009}), and for further evolution of a supernova remnant.

Similar process could be imitated in a laboratory, as a collision of the rapid particle beam with a magnetized gas reservoir filled with a cold neutral low density gas.
Collision of a similar beam may happen with a glass wall, in presence of a magnetic field, producing electromagnetic wave, which may appear in the opposite side, crossing the glass layer.

Transformation of waves with producing of electromagnetic radiation by collision of plasma beam with a magnetic wall, with consideration of different physical processes, was investigated, for example, in the works of  \cite{Usov}  and  \cite{Umeda2011}.

The MHDE equations describe both hydrodynamic and electromagnetic phenomena in the situation when the velocity of the matter are non-relativistic. For description of the behaviour in relativistic jets it is necessary to apply relativistic (Special Relativity) equations of the fluid motion.

However, MHDE equations can be applied to accretion  phenomena in  X-ray pulsars \cite{al1976}, and other flows around neutron stars, white dwarfs, and black holes. They could be applied for description of processes in AGN \cite{lov}, with magnetically driven jets , where the velocity of the fluid and the speed of sound remain non-relativistic. With account of the displacement current, the dispersion equation for the Alfven mode is valid for arbitrary small density and arbitrary large magnetic field. In these limits the Alfven wave is transformed into the electromagnetic one. This property was used for construction of the universal numerical scheme in the frame of MHDE, by McGregor \& Robinson \cite{mcgre}, under the name FMHD.

The summary of our results about wave transformation, originated in regions with high conductivity, during their propagation into regions with low conductivity, across the magnetic field, include three types of behaviour, depending on a magnetization of the flow. 

1. At low magnetization the rapid MHD wave is transforming into the sound wave (Fig. \ref{fig:case1r}). 

2. At intermediate magnetization there is no transformation, the rapid MHD wave is damping, before reaching  of the region of low conductivity (Fig. \ref{fig:case2r}).

3. At high magnetization the rapid MHD wave is transforming into electromagnetic wave (Fig. \ref{fig:case3r}).

The damping of all waves and static perturbations takes place everywhere (Figs. \ref{fig:case1i}, \ref{fig:case2i}, \ref{fig:case3i}).

In the wave propagation along the magnetic field we have the Alfwen wave instead of rapid MHD. Here we have two possibilities.

1. No transformation at low magnetisation with damping of the Alfven wave (Fig. \ref{fig:alf1r}).

2. Transformation of the Alfven wave into electromagnetic at high magnetization (Fig.\ref{fig:alf2r}).

In both cases the waves and static perturbations are damping, according to Figs. \ref{fig:alf1i},\ref{fig:alf2i}. In addition, in both cases non-damping sound waves may be present.

In absence of the magnetic field, only non-damping sound wave, and damping electromagnetic wave may be present. The last ones exist only at sufficiently low conductivity, or sufficiently short  wavelengths (Figs. \ref{fig:em_real},\ref{fig:em_imag}).

\section*{Acknowledgments}
	
This work was supported by the Russian Science Foundation under grant no.23-12-00198. G.S.B.-K. is grateful to Yuri Lyubarsky for stimulating discussion.

\bibliographystyle{unsrt}
 
\bibliography{main}

\begin{thebibliography}{10}

\bibitem{alfven}
H.~Alfven.
\newblock {\em Cosmical Electrodynamics}.
\newblock International Series of Monographs on Physics, Oxford: Clarendon
  Press, 1950.

\bibitem{kulik}
A.G. Kulikovskii and G.A. Lyubimov.
\newblock {\em Magnetohydrodynamics}.
\newblock Addison-Wesley Educational Publishers Inc, 1965.

\bibitem{lleldin}
L.D. Landau, L.P. Pitaevskii, and E.M. Lifshits.
\newblock {\em Electrodynamics of continuous medium}.
\newblock Butterworth-Heinemann, 1984.

\bibitem{sbk21}
A.~A. Soloviev, K.~F. Burdonov, A.~V. Kotov, S.~E. Perevalov, R.~S. Zemskov,
  V.~N. Ginzburg, A.~A. Kochetkov, A.~A. Kuzmin, A.~A. Shaikin, I.~A. Shaikin,
  and et~al.
\newblock Experimental study of the interaction of a laser plasma flow with a
  transverse magnetic field.
\newblock {\em Radiophysics and Quantum Electronics}, 63:876, 2021.

\bibitem{kns23}
S.~V. Korobkov, A.~S. Nikolenko, M.~E. Gushchin, A.~V. Strikovsky, I.~Yu.
  Zudin, N.~A. Aidakina, I.~F. Shaikhislamov, M.~S. Rumenskikh, R.~S. Zemskov,
  and M.~V. Starodubtsev.
\newblock Features of dynamics and instability of plasma jets expanding into an
  external magnetic field in laboratory experiments with compact coaxial plasma
  generators on a large-scale $"$\uppercase{k}rot$"$ stand.
\newblock {\em Astronomy Reports}, 67:93, 2023.

\bibitem{rig}
G.~Rigon, C.~Stoeckl, T.~M. Johnson, J.~Katz, J.~Peebles, and C.~K. Li.
\newblock Development of a platform for experimental and computational studies
  of magnetic and radiative effects on astrophysically-relevant jets at omega.
\newblock {\em arXiv:2401.10779}, 2024.

\bibitem{mcgre}
D.A. McGregor and A.C. Robinson.
\newblock An indirect ale discretization of single fluid plasma without a fast
  magnetosonic time step restriction.
\newblock {\em Computers and Mathematics with Applications}, 78:417, 2019.

\bibitem{bret}
A.~Bret, M.-C. Firpo, and C.~Deutsch.
\newblock Collective electromagnetic modes for beam-plasma interaction in the
  whole $k$ space.
\newblock {\em Physical Review E}, 70(046401), 2004.

\bibitem{liang}
L.~Xiang, B.~Ma, Q.-H. Li, L.~Chen, H.-W. Yu, and D.-J. Wu.
\newblock Effects of displacement current on wave dispersion relation and
  polarization properties in auroral plasmas.
\newblock {\em Research in Astronomy and Astrophysics}, 21 (10):252, 2021.

\bibitem{stas}
K.~Stasiewicz, P.~Bellan, C.~Chaston, C.~Kletzing, R.~Lysak, J.~Maggs,
  O.~Pokhotelov, C.~Seyler, P.~Shukla, L.~Stenflo, A.~Streltsov, and {J. E.}
  Wahlund.
\newblock Small scale alfvenic structure in the aurora.
\newblock {\em Space Science Reviews}, 92:423--533, 2000.

\bibitem{seyfer}
C.~E. Seyler and M.~R. Martin.
\newblock Relaxation model for extended magnetohydrodynamics: Comparison to
  magnetohydrodynamics for dense z-pinches.
\newblock {\em Physics of Plasmas}, 18 (1)(012703), 2011.

\bibitem{buc}
N.~Bucciantini.
\newblock {\em Relativistic hydrodynamics and magneto-hydrodynamics (Lecture
  Notes)}.
\newblock John Wiley and Sons, Inc, 2019.

\bibitem{leis}
{Leismann, T.}, {Antón, L.}, {Aloy, M. A.}, {Müller, E.}, {Martí, J. M.},
  {Miralles, J. A.}, and {Ibáñez, J. M.}
\newblock Relativistic mhd simulations of extragalactic jets.
\newblock {\em Astronomy and Astrophysics}, 436(2):503, 2005.

\bibitem{matt}
{Mattia, G.}, {Del Zanna, L.}, {Bugli, M.}, {Pavan, A.}, {Ciolfi, R.}, {Bodo,
  G.}, and {Mignone, A.}
\newblock Resistive relativistic mhd simulations of astrophysical jets.
\newblock {\em Astronomy and Astrophysics}, 679:A49, 2023.

\bibitem{chap90}
S.~Chapmen and T.G. Cowling.
\newblock {\em Mathematical Theory of Nonuniform Gases}.
\newblock Cambrige, 1952.

\bibitem{brag57}
S.I. Braginskii.
\newblock Transport phenomena in a completely ionized two-temperature plasma.
\newblock {\em Sov. Phys. JETP}, 6:358, 1958.

\bibitem{syr57}
S.I. Syrovatskyi.
\newblock Magnetohydrodynamics.
\newblock {\em Uspekhi Fiz. Nauk}, 62:247, 1957.
\newblock (in Russian); Magnetohydrodynamik. Fortschritte der Physik (1958) 6,
  437 (in German).

\bibitem{wkb}
N.~Fr\"oman and P.Q. Fr\"oman.
\newblock {\em JWKB approximation}.
\newblock N.-Holl. Pub. Comp. Amsterdam, 1965.

\bibitem{bkuniv}
G.S. Bisnovatyi-Kogan.
\newblock Accretion into black hole, and formation of magnetically arrested
  accretion disks.
\newblock {\em Universe}, 5:146, 2019.

\bibitem{rr}
R.~Retes-Romero, Y.~D. Mayya, A.~Luna, and L.~Carrasco.
\newblock Infrared dark clouds and high-mass star formation activity in
  galactic molecular clouds.
\newblock {\em The Astrophysical Journal}, 897(1):53, 2020.

\bibitem{cloud}
{Frieswijk, W. W. F.} and {Shipman, R. F.}
\newblock Searching for dark clouds in the outer galactic plane* - i. a
  statistical approach for identifying extended red(dened) regions in 2mass.
\newblock {\em Astronomy and Astrophysics}, 515:A51, 2010.

\bibitem{gwl}
A.~Giannetti, F.~Wyrowski, S.~Leurini, J.~Urquhart, T.~Csengeri, K.~Menten,
  L.~Bronfman, and F.~Tak.
\newblock Infrared dark clouds on the far side of the galaxy.
\newblock {\em Astronomy and Astrophysics}, 580:L07, 2015.

\bibitem{takiwaki2009}
T.~Takiwaki, K.~Kotake, and K.~Sato.
\newblock Special relativistic simulations of magnetically dominated jets in
  collapsing massive stars.
\newblock {\em The Astrophysical Journal}, 691:1360, 2009.

\bibitem{Usov}
{Usov V.} and {Smolsky M.}
\newblock Wide ultrarelativistic plasma-beam–magnetic-barrier collision.
\newblock {\em Phys Rev E}, 57(2):2267, 1997.

\bibitem{Umeda2011}
Takayuki Umeda.
\newblock {\em Electromagnetic Waves in Plasma. In Wave Propagation. Edited by
  Dr. Andrey Petrin}.
\newblock John Wiley and Sons, Inc, 2011.

\bibitem{al1976}
J.~Arons and S.M. Lea.
\newblock Accretion onto magnetized neutron stars - normal mode analysis of the
  interchange instability at the magnetopause.
\newblock {\em Astrophysical Journal}, 1:792--804, 1976.

\bibitem{lov}
R.V.E. Lovelace.
\newblock Dynamo model of double radio sources.
\newblock {\em Nature}, 262:649, 1976.

\end{thebibliography}

\end{document}